%Paper: gr-qc/9304026
%From: jdbrown@grad00.math.ncsu.edu (John D. Brown)
%Date: Mon, 19 Apr 93 15:44:44 EDT

%%%
%%% To be published in Classical and Quantum Gravity.
%%%
%%% Plain TeX, 40 pages.
%%%
\magnification=\magstep1
\font\twelvebf=cmbx10 at 12pt
\vsize=8.5truein
\hsize=6.0truein
\vskip 0.25in
\baselineskip=15pt plus 0.2pt minus 0.2pt
\parskip=2pt
\parindent=24pt
\hoffset=0.3in
\footline={\ifnum\pageno<2\hfil\else\hss\tenrm\folio\hss\fi}
%%%%%%%%%%%%%%%%%%%%%%
\hyphenation{Euler-ian}
\def\aA{\alpha^{\scriptscriptstyle A}}
\def\bA{\beta_{\scriptscriptstyle A}}
\def\Hfluid{{\cal H}^{\scriptscriptstyle\rm fluid}}
\def\Hifluid{{\cal H}_i^{\scriptscriptstyle\rm fluid}}
\def\sss{\scriptscriptstyle}
\def\F{{\cal F}}
%%%%%%%%%%%%%%%%%%%%
\rightline{November, 1992}
\vskip 1.3in
%\topinsert\vskip 0.3in\endinsert
\centerline{\twelvebf Action functionals for relativistic perfect fluids}
\vskip 0.5in
\centerline{J. David Brown}
\smallskip
\centerline{\it Department of Physics and}
\centerline{\it Department of Mathematics}
\centerline{\it North Carolina State University}
\centerline{\it Raleigh, NC 27695--8202}
\vskip 0.8in
\centerline{ABSTRACT}
\smallskip
\noindent{Action functionals describing relativistic perfect fluids are
presented. Two of these actions apply to fluids whose equations of
state are specified by giving the fluid energy density as a function
of particle number density and entropy per particle. Other actions
apply to fluids whose equations of state are specified in terms of
other choices of dependent and independent fluid variables.
Particular cases include actions for isentropic fluids
and pressureless dust. The canonical Hamiltonian forms of these
actions are derived, symmetries and conserved charges are identified,
and the boundary value and initial value problems are discussed.
As in previous works on perfect fluid actions,
the action functionals considered here depend on certain
Lagrange multipliers and Lagrangian coordinate fields. Particular
attention is paid to the interpretations of these variables and to
their relationships to the physical properties of the fluid. \par}
%\smallskip
%\noindent Classification numbers: 0440, 4775
%\smallskip
%\noindent Short title: Action functionals for relativistic perfect fluids
\vfill\eject
%%%%%%%%%*%%%%%%%%%*%%%%%%%%%*%%%%%%%%%*%%%%%%%%%*%%%%%%%%%*%%%%%%%%%*%%%%%
\noindent{\bf 1. Introduction}
\medskip

This paper contains a description of several action functionals whose extrema
are the classical histories of a relativistic perfect fluid. The possible
benefits of these actions are difficult to judge, but experience has shown
generally that action principles are a powerful and conceptually elegant
means of specifying a dynamical system [1]. They concisely encode the
classical equations of motion, provide a direct and simple way of relating
symmetries and conserved charges, and serve as the starting point for a
canonical Hamiltonian analysis and for Hamilton--Jacobi theory.
For a quantum mechanical description of a system, the existence of an action
functional appears to be a necessity [2].

Perfect fluids are described locally by various thermodynamical variables.
In the notation of Misner, Thorne, and Wheeler [3], they are
$$\eqalignno{ n    &= {\hbox{particle number density}} \ ,&(1.1a)\cr
              \rho &= {\hbox{energy density}} \ ,&(1.1b)\cr
              p    &= {\hbox{pressure}} \ ,&(1.1c)\cr
              T    &= {\hbox{temperature}} \ ,&(1.1d)\cr
              s    &= {\hbox{entropy per particle}} \ .&(1.1e)\cr}$$
These variables are spacetime scalar fields whose values represent
measurements made in the rest frame of the fluid. The fluid motion can be
characterized by its unit four--velocity vector field
$U^a$.\footnote*{Spacetime tensor indices are denoted by $a$, $b$,
{\it etc.\/}, the spacetime metric is $g_{ab}$, and spacetime covariant
derivatives are denoted by a semicolon. The sign conventions of reference
[3] are used thoughout.}

Previous works on relativistic and nonrelativistic
perfect fluid actions [4--13] have revealed a
number of common features that any such action principle must possess. In
particular, it is known that no perfect fluid action can be constructed
solely from the variables (1.1) and $U^a$ unless the variations among those
variables are constrained [5]. Two of the required constraints are particle
number conservation $(nU^a)_{;a} = 0$ and the absence of entropy exchange
between neighboring flow lines $(nsU^a)_{;a}=0$. The remaining constraint is
that the fluid flow lines should be fixed on the boundaries of spacetime.
One way of enforcing this constraint is to vary the flow lines by Lie
dragging along some vector field $\xi^a$, where $\xi^a$ vanishes on the
boundaries. The flow line variation then induces variations in the
four--velocity $U^a$ with the result $\delta U^a = (\delta_b^a + U_b U^a)
{\hbox{\it\$}}_\xi U^b$, where ${\hbox{\it\$}}_\xi$ is the Lie derivative
along $\xi^a$.

There is another method for handling the constraint that the flow lines
should be fixed on the spacetime boundaries, suggested by the work of Lin
[9, 10]. Instead of the four--velocity $U^a$, the history of the fluid is
characterized by a set of spacetime scalar fields $\aA$, $A=1$, $2$, $3$,
that are interpreted as Lagrangian coordinates for the fluid. That is,
$\aA(x)$ serve as labels for the fluid, specifying which flow line passes
through a given spacetime point $x$. A set of Lagrangian coordinates can be
generated by choosing an arbitrary spacelike hypersurface and a coordinate
system $\aA$ on that surface. Then each flow line is labeled by the
coordinate value of the point where it intersects the hypersurface. By
building an action functional using the Lagrangian coordinates $\aA$, the
fluid flow lines are held fixed on the spacetime boundaries by simply fixing
$\aA$ on the boundaries.

The particle number and entropy exchange constraints can be incorporated
directly into the action via Lagrange multipliers. An action $S$ is presented
in section 2 that makes use of such Lagrange multipliers along with the
Lagrangian coordinates $\aA$ to enforce the constraints. This action
describes a perfect fluid whose equation of state is specified by giving
the energy density $\rho$ as a function of number density $n$ and entropy
per particle $s$. The action $S$ incorporates various features found in
the action functionals discussed in references [5, 8, 14--16], but to
my knowledge it has not been written previously in precisely the form
discussed here. In sections 2 and 3 the Lagrangian and Hamiltonian forms
of the action $S$ are developed, the Lagrangian and Hamiltonian equations
of motion are displayed explicitly, global symmetries and conserved charges
are analyzed, and the initial and boundary value problems are discussed.

The use of Lagrangian coordinates $\aA$ and Lagrange multipliers is
sometimes criticized on the grounds that these extra variables are
``unphysical". One of the results of the present work is to show that
the values of each of these variables are determined to within a global
symmetry transformation by the physical properties of the fluid. The
situation here is formally analogous to
that of a free nonrelativistic point particle moving in flat space with
cartesian coordinates. In that case, Newton's law is $\dot{\vec v} = 0$,
where $\vec v$ is the particle's velocity and the dot denotes a time
derivative. An action functional that yields this equation of motion is
$\int dt (\vec v\cdot\vec v/2)$, where the variations in $\vec v $ are
constrained to those having the form $\delta\vec v = \dot{\vec\xi}$. Here,
$\vec\xi$ is an infinitesimal vector that vanishes at the initial and final
times. An unconstrained action is obtained by introducing a ``Lagrangian"
coordinate $\vec x$ and constructing the velocity from $\vec v =
\dot{\vec x}$. The new variable $\vec x$ is just the particle's coordinate
location. The value of $\vec x$ at any given time specifies the physical
location of the particle to within a symmetry translation or rotation of
the cartesian coordinates. Similarly, the values of the Lagrangian
coordinates $\aA$ specify the physical location of the fluid flow lines to
within a symmetry transformation that amounts to a change of Lagrangian
coordinates. The values of the Lagrange multipliers that appear in the fluid
action $S$ are also determined to within a global symmetry transformation
by certain physical properties of the fluid. (The action $S$ is the analogue
of the free particle action $\int dt (\vec v\cdot \dot{\vec x} -
\vec v\cdot\vec v/2 )$ in the sense that it is a functional of both the
four--velocity $U^a$ and the Lagrangian coordinates $\aA$.)

One of the fluid equations of motion obtained from the action $S$ relates
the fluid four--velocity $U_a$ to the
Lagrange multipliers, Lagrangian coordinates $\aA$, and their gradients.
Expressions of this type are often called velocity--potential
representations or Clebsch [29] representations of the four--velocity. There
have been numerous discussions in the literature concerning both the number
of scalar fields that are mathematically required for such a representation,
and the number of scalar fields that are physically natural for such a
representation [4, 5, 10, 12, 15]. Section 4 contains a further discussion
of these issues.

The Hamiltonian form of the perfect fluid action $S$ is a functional of the
fluid number density, the fluid entropy density, the Lagrangian coordinates
$\aA$, and their canonical conjugates. In this case the number and entropy
densities are Eulerian, that is, their values correspond to measurements
made by observers at rest in space. The Eulerian densities are related
to the Lagrangian (or comoving) densities $n$ and $ns$ by a kinematical
``gamma" or boost factor that in turn is determined by the local spatial
velocity of the fluid. The fluid contributions to the Hamiltonian and
momentum constraints are just the appropriate projections of the perfect
fluid stress--energy--momentum tensor [17], and involve the spatial
components $U_i$ of the fluid four--velocity. These components $U_i$ are
explicitly expressed in terms of the canonical fluid variables and the
Lagrangian particle number density $n$. The Lagrangian number density $n$ is
itself a function of the canonical variables as implicitly determined by
the equation relating the Lagrangian and Eulerian number densities.

A Hamiltonian formulation of perfect fluids that makes use of only
``physical" fluid variables has been developed (see [18] and references
therein) based on the Lie--Poisson brackets. However, Lie--Poisson brackets
are not canonical, so there is no underlying symplectic structure and the
usual ``$\int(p\dot q - H)$" form of the action does not apply. On the other
hand, the Lie--Poisson structure can be derived by a so--called Lagrangian
to Eulerian map starting from a certain canonical Hamiltonian structure that
involves the Lagrangian coordinates [18]. This canonical Hamiltonian
formulation of perfect fluids is derived in section 5, starting from an
associated action $\bar S$ that is a functional of the Lagrangian
coordinates $\aA$ only. In this action, the particle number and entropy
exchange constraints are enforced by restricting $nU^a$ and $s$ to depend on
spacetime only through certain combinations of the Lagangian
coordinates. In particular, $s$ is given by a function $s(\alpha)$, and
$nU^a$ is given by $-\eta_{123}\epsilon^{abcd}{\alpha^1}_{,b}
{\alpha^2}_{,c}{\alpha^3}_{,d}$, where $\eta_{123}$ is a function of $\aA$.
($\epsilon^{abcd}$ is obtained by raising indices on the totally
antisymmetric spacetime volume form $\epsilon_{abcd}$.) The action $\bar S$,
like $S$, describes a perfect fluid whose equation of state is specified by
giving the energy density $\rho$ as a function of number density $n$ and
entropy per particle $s$.

In addition to the local thermodynamical variables (1.1), it is also
convenient to define [3]
$$\eqalignno{\mu &:= {\rho + p\over n} = {\hbox{chemical potential}}
                    \ ,&(1.2a)\cr
              a &:= {\rho\over n} - Ts = {\hbox{physical free energy}}
                    \ ,&(1.2b)\cr
              f &:= {\rho + p\over n} - Ts = {\hbox{chemical free energy}}
                    \ .&(1.2c)\cr}$$
The chemical potential $\mu$ is the energy per particle required to inject
a small amount of fluid into a fluid sample, keeping the sample volume and
the entropy per particle $s$ constant [3]. Similarly, the physical free
energy $a$ is the injection energy at constant number density and constant
total entropy, while the chemical free energy $f$ is the injection energy at
constant volume and constant total entropy. The thermodynamical variables
(1.1--2) are related by the local expression of the first law of
thermodynamics, namely [3]
$$d\rho = \mu\,dn + nT\,ds \ .\eqno(1.3)$$
This relationship shows that an equation of state for a perfect fluid can
be specified by giving the function $\rho(n,s)$, the energy density as a
function of number density and entropy per particle. The first law also can
be written as
$$dp = n\,d\mu - nT\,ds \eqno(1.4)$$
which naturally suggests an equation of state of the form $p(\mu,s)$.
Another possibility is
$$d(na) = f\,dn - ns\,dT \eqno(1.5)$$
which suggests an equation of state $a(n,T)$. In section 6, various action
functionals are presented for perfect fluids specified by the equations of
state $p(\mu,s)$ and $a(n,T)$. As special cases,  actions for isentropic
fluids and pressureless dust are obtained.

The perfect fluid action functionals developed by Taub [6, 7] and Schutz
[15, 19, 20] have been applied to the analysis of the stability of stellar
models [7, 21--23], and it is hoped that the actions discussed here can be
put to similar use. Schutz's action also has been used to quantize the
combined gravity--fluid system [24, 25], where one of the Lagrange
multipliers is chosen as a time coordinate. Along these same lines, the
action for a pressureless dust, discussed in section 6, has been used in an
investigation of the problem of time in canonical quantum gravity [26].
Another application of perfect fluid actions is found in the work of Gibbons
and Hawking [27] (also see [28] and references therein). There, the action
is used to obtain a thermodynamical potential that characterizes the
global thermal properties of a gravitating fluid system.

%%%%%%%%%*%%%%%%%%%*%%%%%%%%%*%%%%%%%%%*%%%%%%%%%*%%%%%%%%%*%%%%%%%%%*%%%%%
\bigskip
\noindent{\bf 2. Action $S$ with equation of state $\rho(n,s)$}
\medskip
\noindent{\it 2.1 Action and equations of motion}
\medskip
By definition, the stress--energy--momentum tensor for a perfect fluid has
the form
$$T^{ab} = \rho U^a U^b + p( g^{ab} + U^a U^b) \ ,\eqno(2.1)$$
where $U^a$ is the unit four--velocity of the fluid. The equations of motion
for a perfect fluid consist of the stress tensor equation of motion, namely
$T^{ab}{}_{;b}=0$, and the equation $(n U^a)_{;a}=0$ expressing conservation
of particle number [3]. The action functional presented in this section
incorporates the stress tensor (2.1), the required equations of motion, and
the first law of thermodynamics (1.3) for a perfect fluid with equation of
state $\rho(n,s)$.

The perfect fluid action $S$ is a functional of a spacetime contravariant
vector density $J^a$ that is interpreted as the densitized particle number
flux vector $\sqrt{-g}nU^a$. That is, the fluid four--velocity is defined
by
$$U^a := J^a/|J| \ ,\eqno(2.2)$$
where
$$|J| := \sqrt{-J^a g_{ab} J^b} \eqno(2.3)$$
is the magnitude of $J^a$, and the particle number density is given by
$$n := |J|/\sqrt{-g} \ .\eqno(2.4)$$
This action is also a functional of the spacetime metric $g_{ab}$, the
entropy per particle $s$, the Lagrangian coordinates $\aA$, and spacetime
scalars denoted by $\varphi$, $\theta$, and $\bA$. (The indices $A$, $B$
take the values $1$, $2$, $3$.) In terms of an arbitrary equation of state
$\rho(n,s)$, the action reads
$$\eqalignno{ S[g_{ab}, J^a, \varphi, \theta, s, \aA, \bA] = \int d^4x
    \Bigl\{ & -\sqrt{-g}\, \rho(|J|/\sqrt{-g}, s) \cr
    &\qquad\quad + J^a (\varphi_{,a} +  s \theta_{,a} + \bA{\aA}_{,a})
       \Bigr\} \ .\qquad &(2.5)\cr}$$
The stress--energy--momentum tensor derived from this action is
$$T^{ab} := {2\over\sqrt{-g}} {\delta S\over\delta g_{ab}} =
    \rho U^a U^b + \biggl( n{\partial\rho\over\partial n} - \rho\biggr)
     \Bigl( g^{ab}  + U^a U^b\Bigr) \ ,\eqno(2.6)$$
and has the perfect fluid form (2.1) with energy density $\rho$ and pressure
defined by
$$p := n{\partial\rho\over\partial n} - \rho \ .\eqno(2.7)$$
This expression for pressure agrees with the relationship implied by the
first law of thermodynamics (1.3).

The fluid equations of motion derived from the action (2.5) are as follows:
$$\eqalignno{ 0 =& {\delta S\over\delta J^a} = \mu U_a + \varphi_{,a} +
                    s\theta_{,a}    + \bA{\aA}_{,a}  \ ,&(2.8a)\cr
              0 =& {\delta S\over\delta\varphi} = -J^a{}_{,a} \ ,&(2.8b)\cr
              0 =& {\delta S\over\delta\theta} = -(sJ^a)_{,a} \ ,&(2.8c)\cr
              0 =& {\delta S\over\delta s} = -\sqrt{-g} {\partial\rho\over
                      \partial s}     + \theta_{,a} J^a  \ ,&(2.8d)\cr
              0 =& {\delta S\over\delta\aA} = - (\bA J^a)_{,a} \ ,&(2.8e)\cr
              0 =& {\delta S\over\delta\bA} = {\aA}_{,a} J^a \ .&(2.8f)\cr}$$
Solutions to these equations extremize the action $S$ with respect to
variations that leave $\varphi$, $\theta$, and $\aA$ fixed on the spacetime
boundaries. According to equations (2.8b,c), the fields $\varphi$ and
$\theta$ serve as Lagrange multipliers for the particle number conservation
constraint $J^a{}_{,a}=0$ and the entropy exchange constraint $(sJ^a)_{,a}=0$,
respectively. Equation (2.8f)
shows that $\bA$ are Lagrange multipliers for the constraints ${\aA}_{,a}J^a
= 0$ that restrict the fluid four--velocity vector to be directed along the
flow lines $\aA = {\rm constant}$.

As discussed in the introduction, the
fields $\aA(x)$ are interpreted as Lagrangian coordinates for the fluid and
serve as labels that specify which flow line passes through a given spacetime
point $x$. A set of Lagrangian coordinates can be generated by choosing an
arbitrary spacelike hypersurface and specifying a coordinate system $\aA$
on that surface. Then each flow line is labeled by the coordinate value of
the point where it intersects the hypersurface. It is useful to view this
arbitrary spacelike hypersurface as an abstract ``fluid space" whose
points represent the fluid flow lines, and to view the Lagrangian
coordinates $\aA$ as a coordinate system on the fluid space. (The
concept of ``fluid space" is also discussed in reference [14], where it is
called the ``matter space".) So
conceptually, the fluid space is the space of Lagrangian coordinate labels,
and as a manifold it is isomorphic to any spacelike hypersurface. It may be
impossible to cover the fluid space with a single coordinate chart, so the
Lagrangian coordinates $\aA$ generally must be defined in open subsets of
the fluid space. With the above interpretation, it is assumed that the
fields $\aA$ constitute a good set of Lagrangian coordinates, in the sense
that within the appropriate open subsets each flow line carries a unique
label $\aA$ and the gradient ${\aA}_{,a}$ is nonvanishing.

The significance of the scalar field $\theta$ is revealed by comparing the
equation of motion (2.8d) with the first law of thermodynamics (1.3). This
leads to the identification
$$T = \theta_{,a}U^a = {1\over n}{\partial\rho\over\partial s} \eqno(2.9)$$
for the fluid temperature $T$, and shows that $\theta$ is the ``thermasy"
discussed by van Dantzig [30]. Kijowski {\it et al.\/} [14] have proposed a
physical interpretation for thermasy, relating it to the difference between
proper time along the fluid flow lines and proper time along the history of
a typical fluid particle that executes chaotic motion about the flow lines.
The field $\varphi$ plays a role that is mathematically analogous to $\theta$,
as seen by contracting equation (2.8a) with $U^a$ and using equations (2.8f)
and (2.9) to obtain
$$f = \varphi_{,a}U^a \ .\eqno(2.10)$$
Thus, $\theta$ is a ``potential" for the temperature $T$ and $\varphi$ is
a ``potential" for the chemical free energy $f$. The interpretations of the
fields $\theta$, $\varphi$, and $\bA$ are discussed further in sections 3
and 4.

As stated above, the perfect fluid equations of motion (2.8) should imply
particle number conservation and the stress tensor equation of motion
$T^{ab}{}_{;b}=0$. The first of these, particle number conservation, is
expressed explicitly in equation (2.8b).  For the stress tensor equation
of motion, first consider its projection along the fluid flow lines. This
yields [3]
$$ 0 = U_a T^{ab}{}_{;b}  =  - {\partial\rho\over\partial s}  s_{,a} U^a
         \ ,\eqno(2.11)$$
where particle number conservation has been used.  This equation combines
particle number conservation (2.8b) and the entropy exchange constraint
(2.8c), and is indeed implied by the equations of motion. This result shows
that the fluid flow is locally adiabatic, that is, the entropy per particle
along the fluid flow lines is conserved.

The projection of the stress tensor equation of motion orthogonal to the
flow lines gives the Euler equation, relating the fluid acceleration to the
gradient of pressure [3]:
$$\eqalignno{ 0 &= \bigl(g_{ab} + U_aU_b\bigr) T^{bc}{}_{;c} \cr
                &= (\rho + p) U_{a;b} U^b + \bigl(\delta_a^b + U_aU^b\bigr)
                  p_{,b}    \ .&(2.12)\cr}$$
Using expression (2.7) for $p$, the Euler equation can be written as
$$\eqalignno{ 0 &= \mu U_{a;b} U^b + \bigl(\delta_a^b + U_aU^b\bigr)
                      {1\over n}p_{,b} \cr
                &= 2(\mu U_{{\sss [}a} )_{;b{\sss ]}} U^b
                   + (\mu U_b)_{;a}U^b - \biggl({\partial\rho\over\partial
                   n}\biggr)_{;b}    U_aU^b \cr
                 &\qquad + \bigl(\delta_a^b + U_aU^b\bigr) {1\over n}\biggl(
                   n {\partial\rho\over\partial n} - \rho\biggr)_{;b} \cr
                &= 2(\mu U_{{\sss [}a} )_{;b{\sss ]}} U^b
                   - \bigl(\delta_a^b + U_aU^b\bigr){1\over n}
                   {\partial\rho\over\partial s}  s_{,b}  \ ,&(2.13)\cr}$$
where the square brackets denote antisymmetrization. Now combine
equation (2.13) with the entropy conservation equation (2.11) to obtain
$$ 2 V_{{\sss [}a;b{\sss ]}} U^b = Ts_{,a} \ ,\eqno(2.14)$$
where
$$ V_a := \mu U_a  \eqno(2.15)$$
and the definition (2.9) for temperature $T$ has been used. The vector
$V_a$, sometimes called the Taub current, plays an important role in
relativistic fluid dynamics, especially in the description of circulation
and vorticity [31]. From the interpretation of $\mu$ as an injection
energy, $V^a$ is identified with the four--momentum per particle of a small
amount of fluid to be injected in a larger sample of fluid without changing
the total fluid volume or the entropy per particle.

It remains to be verified that the equations of motion (2.8) imply the Euler
equation (2.12). For this purpose, it suffices to show that the equations
of motion imply equation (2.14), since that equation and the Euler equation
are related by use of the equations of motion. Accordingly, compute
$2V_{{\sss [}a;b{\sss ]}} U^b $ using the expression for $V_a$ from equation
(2.8a) to obtain
$$ 2V_{{\sss [}a;b{\sss ]}} U^b = - 2( \varphi_{,{\sss [}a} + s
     \theta_{,{\sss [}a} +  \bA{\aA}_{,{\sss [}a} )_{;b{\sss ]}} U^b
      \ .\eqno(2.16)$$
The equations of motion imply that the terms involving  gradients of $s$,
$\aA$ and $\bA$ along the flow lines all vanish. Only the term
$s_{,a}\theta_{,b}U^b$ remains on the right--hand--side of equation (2.16).
According to equation (2.9), this term equals $Ts_{,a}$ so that the
equations of motion indeed imply equation (2.14).  The conclusion is that
the equations of motion (2.8) derived from the action (2.5) imply the
relevant perfect fluid equations of motion, including the Euler equation.

The ``on shell" action, that is, the value of the action (2.5) when the
equations of motion (2.8) hold, is just the proper volume integral of the
pressure:
$$S({\hbox{on shell}}) = \int d^4x \sqrt{-g}\, p \ .\eqno(2.17)$$
Of course, the addition of surface integrals to the action will change its
on--shell value without affecting the equations of motion. In particular,
the surface integral $-\int d^4x (\varphi J^a)_{,a}$ can be added to $S$,
which amounts to replacing the term $\varphi_{,a} J^a$ by the term $-\varphi
J^a{}_{,a}$. Likewise, adding the surface integral $-\int d^4x (\theta s
J^a)_{,a}$ to $S$ results in replacing $ s\theta_{,a} J^a$ by
$-\theta (sJ^a)_{,a}$. These surface integrals have on--shell values
$-\int d^4x \sqrt{-g} nf$ and $-\int d^4x \sqrt{-g} nTs$, respectively.
Thus, by adding these surface integrals in various combinations, action
functionals can be obtained whose on--shell values are, for example, the
proper volume integrals of $-\rho$ or $-na$.

There are a number of alternative perfect fluid actions that differ
from the action (2.5) in the placement of derivatives in each term. In some
cases, such as changing $\varphi_{,a}J^a$ to $-\varphi {J^a}_{,a}$, this
amounts to the simple addition to the action of a boundary term as discussed
above. A valid action functional is also obtained by replacing
$s\theta_{,a}J^a$ with $-\theta s_{,a} J^a$. This change is brought about
by defining $\varphi := \varphi' - s\theta$ then dropping the prime on the
new field ($\varphi'\to\varphi$). In place of equation (2.10), $\varphi$
then satisfies $\mu = \varphi_{,a}U^a$ and is a ``potential" for the
chemical potential. Nevertheless, the resulting action yields the required
equations of motion and stress tensor. The action (2.5) has an advantage
over its alternatives in that the momenta conjugate to $\varphi$ and $\theta$
that naturally follow from that action have clear physical interpretations
as the particle number density and entropy density seen by (Eulerian)
observers at rest in space.

%%%%%%%%%*%%%%%%%%%*%%%%%%%%%*%%%%%%%%%*%%%%%%%%%*%%%%%%%%%*%%%%%%%%%*%%%%%
\bigskip
\noindent{\it 2.2 Symmetries of the action $S$}
\medskip
The perfect fluid action (2.5) is invariant under the infinitesimal
transformations\footnote*{These symmetries are derived in the Appendix.}
$$\eqalignno{ \delta\varphi &= \epsilon F - \epsilon s(\partial F/\partial
                      s) - \epsilon\bA(\partial F/\partial\bA) &(2.18a)\cr
              \delta\theta  &= \epsilon(\partial F/\partial s) &(2.18b)\cr
              \delta s      &= 0 &(2.18c)\cr
              \delta\aA     &= \epsilon(\partial F/\partial\bA) &(2.18d)\cr
              \delta\bA     &= - \epsilon(\partial F/\partial\aA)
                      \ ,&(2.18e)\cr}$$
along with $\delta J^a =0$ and $\delta g_{ab}=0$, where $\epsilon$ is an
infinitesimal parameter and $F$ is an arbitrary function of the Lagrangian
coordinates $\aA$, the entropy
per particle $s$, and $\bA$. These invariances, one for each function
$F(\alpha,\beta,s)$, constitute global symmetries of the theory, symmetries
in which the field transformations are fixed for all time. Such a
symmetry is to be distinguished from a gauge symmetry, in which the
transformation can vary from one instant of time to the next. Observe that
the particular combination $\varphi_{,a} +  s\theta_{,a}  + \bA{\aA}_{,a}$
that appears in the action (2.5), and equals $-\mu U_a$ when the equation
of motion (2.8a) holds, is invariant under the transformations (2.18).
Also note that the function $F(\alpha,\beta,s)$ has vanishing gradient
$F_{,a}U^a = 0$ along the fluid flow lines, according to the equations
of motion (2.8). Thus, the symmetries (2.18) produce variations in
$\varphi$, $\theta$, $\aA$, and $\bA$ that are constant along the flow
lines. Such variations do not affect the temperature $T = \theta_{,a}U^a$,
chemical free energy $f = \varphi_{,a}U^a$, or the constancy of $\aA$ and
$\bA$ along the flow lines.

The physical meaning of the symmetries (2.18) can be explored by considering
special choices for the function $F(\alpha,\beta,s)$. If $F$ is a function
of $\aA$ only, the transformations (2.18) reduce to
$$\eqalignno{\delta\varphi &= \epsilon F  &(2.19a)\cr
         \delta\bA &= - \epsilon (\partial F/\partial\aA) \ ,&(2.19b)\cr}$$
with all other fields unchanged. Since the Lagrangian coordinates $\aA$
uniquely label the flow lines, at least within open subsets of the fluid
space, the change (2.19a) in $\varphi$ along one flow line is independent of
the changes in $\varphi$ along the other flow lines (subject to the
requirement of continuity in $\varphi$). Therefore, the symmetries (2.19)
can be understood as arising from the freedom to shift the value of
$\varphi$ along each flow line by a constant amount. In particular, any
solution to the fluid equations of motion can be transformed via equations
(2.19) into a solution with, say, $\varphi=0$ on any given spacelike
hypersurface.

If the function $F$ has the form $s\,\tilde F(\alpha)$ for some function
$\tilde F$ of $\aA$, the transformations (2.18) reduce to
$$\eqalignno{\delta\theta &= \epsilon\tilde F  &(2.20a)\cr
           \delta\bA     &= - \epsilon s(\partial\tilde F/\partial\aA)
             \ ,&(2.20b)\cr}$$
with the other fields unchanged. These symmetries can be used to shift the
value of the thermasy $\theta$ along each flow line by a constant amount.
Thus, any solution of the equations of motion can be transformed into
a solution with $\theta=0$ on any given spacelike hypersurface.

Now consider functions $F$ of the form $\bA\bar F^{{\sss A}}(\alpha)$ where
$\bar F^{{\sss A}}$ is a set of functions of $\aA$. In this
case, the transformations (2.18) reduce to
$$\eqalignno{\delta\aA &= \epsilon\bar F^{{\sss A}} &(2.21a)\cr
             \delta\bA &= -\epsilon\beta_{{\sss B}} (\partial
                \bar F^{{\sss B}} /\partial\aA) \ .&(2.21b)\cr}$$
These symmetry transformations describe changes of coordinates $\aA$ in the
fluid space, where $\bar F^{{\sss A}}$ is viewed as a vector
in the fluid space. Moreover, observe that the equations of motion (2.8b,e)
imply the constancy of $\bA$ along the fluid flow lines, so that $\bA$ can be
expressed as a function of the Lagrangian coordinates $\aA$. Then according
to its transformation (2.21b) under changes of coordinates (2.21a), $\bA$
can be viewed as the components of the covariant vector (or one--form)
$\beta := \bA d\aA$ in the fluid space.

The conserved Noether currents associated with the symmetries (2.18) are
obtained from the general variation of the action (2.5), which is
$$\eqalignno{ \delta S =& \,({\hbox{terms giving the equations of
                  motion}}) \cr
               &+ \int d^4x (J^a\delta\varphi + sJ^a\delta\theta  +
                 \bA J^a\delta\aA)_{,a}  \ .&(2.22)\cr}$$
The action $S$ is invariant under the symmetry transformations (2.18), so
when the equations of motion hold, $0=\delta S$ becomes
$$0 = \int d^4x (F J^a)_{,a} \ .\eqno(2.23)$$
Therefore $FJ^a$ are the conserved currents that satisfy $(FJ^a)_{,a}=0$
by virtue of the equations of motion (2.8). If space is closed, or if
suitable boundary conditions are imposed at spatial infinity, the volume
integral (2.23) can be written as a surface integral $Q[F]$ evaluated on a
final spacelike hypersurface, minus the same surface integral evaluated on
an initial spacelike hypersurface. That surface integral is the Noether
charge
$$Q[F] = \int_\Sigma d^3x \sqrt{h}\, n(-n_aU^a)F(\alpha,\beta,s)
          \ ,\eqno(2.24)$$
and equation (2.23) just expresses the conservation of $Q[F]$, that is, the
independence of $Q[F]$ on the choice of hypersurface $\Sigma$. In equation
(2.24), $h$ is the determinant of the three--metric on $\Sigma$, and $n_a$
is the future pointing unit normal to $\Sigma$.

Insight into the meaning of the conserved charges $Q[F]$ is obtained by
considering again a special choice for $F(\alpha,\beta,s)$. Let $F$ be a
function of $\aA$ only, and in particular choose $F$ equal
to unity for all $\aA$ in some ball $B$ in the fluid space, and equal to
zero outside $B$. Then the charge (2.24) becomes a proper volume integral
over the subspace of the hypersurface $\Sigma$ that contains flow lines
in $B$, with integrand $n(-n_aU^a)$. The factor $-n_aU^a$ is the
relativistic ``gamma factor" characterizing a boost from the Lagrangian
observers with four--velocity $U^a$ to the Eulerian observers with
four--velocity $n^a$ [32]. Thus, $n(-n_aU^a)$ is the particle number
density as seen by the Eulerian observers. Then conservation of the
charge (2.24) expresses the conservation of particle number within a
flow tube defined by the bundle of flow lines contained in the ball $B$.

For general functions $F(\alpha,\beta,s)$, the conserved charges (2.24)
can be given a physical interpretation in terms of the case discussed
above. Since $\bA$ and $s$ are constant
along the flow lines, and because $\aA$ are unique flow line labels, the
fields $\bA$ and $s$ can be expressed as functions of the spacetime point
$x$ through the combination $\aA(x)$. Thus, a general function $F$ can be
viewed as having functional dependence $F(\alpha,\beta(\alpha),s(\alpha))$.
Then the charge $Q[F]$ is the number of particles contained in a
distribution of flow lines, with each flow line weighted by the distribution
function $F(\alpha,\beta(\alpha),s(\alpha))$. For example, choose $F$ equal
to $s\tilde F$, where $\tilde F$ equals unity for all $\aA$ in a ball $B$,
and zero outside $B$. In this case the conserved charge $Q[F]$ equals the
number of particles contained in the flow lines included in $B$, weighted
by the entropy per particle $s$. Conservation of $Q[F]$ expresses the
conservation of total entropy within the flow tube defined by the bundle
of flow lines contained in $B$.

The conserved charges (2.24) must be distinguished from quantities that
are conserved in a more restricted sense, such as the fluid circulation
$C$. Circulation is defined as the integral of the Taub current one--form
$V_a =\mu U_a$ around a spacelike loop [31]. $C$ is conserved in the
sense that its value is a constant for particular families of loops,
namely those obtained by evolving an initial loop along the fluid flow
lines by an amount proportional to the thermasy $\theta$ [33]. (If the
loop is evolved along the flow lines by an amount proportional to proper
time, changes in $C$ are determined by the gradient of the entropy per
particle [31].) On the other hand, a conserved charge $Q[F]$ has the same
value for any spacelike hypersurface, not just for a particular family of
hypersurfaces. See reference [34] for a discussion of the various types
of conservation laws that arise in fluid mechanics.

%%%%%%%%%*%%%%%%%%%*%%%%%%%%%*%%%%%%%%%*%%%%%%%%%*%%%%%%%%%*%%%%%%%%%*%%%%%
\bigskip
\noindent{\bf 3. Hamiltonian form of the action $S$}
\medskip
\noindent{\it 3.1 Action and Hamiltonian}
\medskip
The perfect fluid action (2.5) can be written in Hamiltonian form by
introducing first a space--time decomposition of the fields. Accordingly,
let $t$ denote a scalar function on spacetime, whose gradient is nonzero and
(future pointing) timelike. The $t={\rm constant}$ surfaces foliate spacetime
into spacelike hypersurfaces with unit normal $n_a = -N t_{,a}$, where $N :=
(-t_{,a} g^{ab} t_{,b})^{-1/2}$ defines the lapse function. Also introduce a
time vector field $t^a$ such that $t_{,a} t^a = 1$ and the projection tensor
$h^a_b := \delta^a_b + n^an_b$ onto the leaves of the foliation. With the
shift vector defined by $N^a := h^a_b t^b$, the time vector field can be
written as $t^a = Nn^a + N^a$.

Using the above relationships, the vector density $J^a$ is decomposed as
$$\eqalignno{ J^a &= h^a_bJ^b - n^an_bJ^b \cr
                  &= (hJ)^a + (t^a - N^a) \Pi \ ,&(3.1)\cr}$$
where the definitions $(hJ)^a := h^a_b J^b$ and $\Pi := t_{,a}J^a$ are used.
Because  $J^a = \sqrt{-g}nU^a$ is the densitized particle number flux vector,
this latter definition becomes
$$\Pi = \sqrt{h} n(-n_aU^a) \ ,\eqno(3.2)$$
where $\sqrt{-g}=N\sqrt{h}$ has been used. Recall that $-n_aU^a$ is the
relativistic ``gamma factor" characterizing a boost from the Lagrangian
observers with four--velocity $U^a$ to the Eulerian observers with
four--velocity $n^a$. Then $\Pi$ is recognized as the spatially densitized
Eulerian particle number density; that is, $\Pi/\sqrt{h}$ is the number
density of fluid particles as seen by the Eulerian observers who are at rest
in the $t={\rm constant}$ hypersurfaces.

With the decomposition (3.1), the action (2.5) becomes
$$\eqalignno{ S =& \int d^4x \Bigl\{-\sqrt{-g}\rho(|J|/\sqrt{-g}, s)
                  +\Pi t^a(\varphi_{,a}
                  + s\theta_{,a} + \bA{\aA}_{,a}) \cr
     &\qquad\qquad\qquad\quad +\bigl( (hJ)^a - \Pi N^a\bigr) (\varphi_{,a}
                  +s \theta_{,a} + \bA{\aA}_{,a}) \Bigr\} \ .&(3.3)\cr}$$
Now tie the spacetime coordinates to the foliation by choosing $t$ as the
``timelike" coordinate, and choosing spatial coordinates such that
$(\partial/\partial t)^a = t^a$. In these coordinates, the spacetime metric
has the ADM form [35]
$$g_{ab} dx^adx^b = -N^2 dt^2 + h_{ij} (dx^i + N^i dt)(dx^j + N^j dt)
             \ ,\eqno(3.4)$$
where $x^i$ are the spatial coordinates on the $t={\rm constant}$
hypersurfaces, and $h_{ij}$ are the spatial components of the spacetime
tensor $h_{ab}$. Note that derivatives along $t^a$ are just ordinary $t$
derivatives, and will be denoted by a dot. Since  the contravariant tensors
$N^a$ and $(hJ)^a$ have vanishing $t$--components, the fluid action (3.3)
can be written as
$$\eqalignno{ S =& \int dt\,d^3x \Bigl\{ \Pi (\dot\varphi  +s \dot\theta
         + \bA\dot\aA)   - N\sqrt{h}\,\rho(|J|/N\sqrt{h},s) \cr
      &\qquad\qquad\qquad\quad + \bigl( (hJ)^i - \Pi N^i\bigr) (\varphi_{,i}
        +s\theta_{,i}     +\bA {\aA}_{,i} \Bigr\} \ ,&(3.5)\cr}$$
where
$$|J| = \Bigl[ (N\Pi)^2 - (hJ)^i(hJ)_i \Bigr]^{1/2} \ ,\eqno(3.6)$$
is the magnitude of $J^a$. (Spatial indices are raised and lowered by the
spatial metric $h_{ij}$ and its inverse $h^{ij}$.)

Since the fields with a time derivative appear linearly in the action (3.5),
it is clear that $\Pi$ is canonically conjugate to $\varphi$, $\Pi s$ is
canonically conjugate to $\theta$, and $\Pi\bA$ is canonically conjugate to
$\aA$. Only the fields $(hJ)^i$ have no canonical counterpart. The action can
be cast into canonical form with respect to the variables $(hJ)^i$ by
following the standard analysis due to Dirac [36]. The momenta conjugate to
$(hJ)^i$ are constrained to vanish, and the preservation in time of these
primary constraints is equivalent to the equation of motion produced by
varying the action with respect to $(hJ)^i$; this is just the projection
of the equation of motion (2.8a) onto the spacelike hypersurfaces, namely
$$U_i = -(\varphi_{,i} +s\theta_{,i} + \bA{\aA}_{,i})/\mu \ .\eqno(3.7)$$
The complete set of constraints is second class, and their elimination is
equivalent to replacing $(hJ)^i$ in the action (3.5) by the solution to the
equation of motion (3.7).

Equation (3.7) can be solved implicitly for $(hJ)^i$ as a function of
$\Pi$, $\varphi$, $\theta$, $s$, $\bA$, $\aA$, and $h_{ij}$ in the
following way. Observe that $(hJ)^i$ appears on the left--hand--side of
that equation through the combination  $U_i = (hJ)_i/|J|$, with $|J|$
given by equation (3.6). The equation $U_i = (hJ)_i/|J|$ can be solved
for $(hJ)^i$ as a function of $U_i$, with the result
$$(hJ)^i = {N\Pi \over \sqrt{1 + U^jU_j}} U^i \ .\eqno(3.8)$$
Now consider $U_i$ as shorthand notation for the right--hand--side of
equation (3.7). This expression for $U_i$ depends on the chemical potential
$\mu = \partial\rho/\partial n$, which is a function of $n$ and $s$ as
determined by the equation of state $\rho(n,s)$. In turn, $n$ is
determined implicitly as a function of $\Pi$, $\varphi$, $\theta$, $s$,
$\bA$, $\aA$, and $h_{ij}$ by the equation
$$\Pi = \sqrt{h} n(1+U_i h^{ij} U_j)^{1/2} \ .\eqno(3.9)$$
This expression (3.9) relates the Lagrangian number density $n$ to the
Eulerian number density $\Pi/\sqrt{h}$, and is obtained from equation (3.2)
with the gamma factor written as $-n_a U^a = (1+U_i h^{ij} U_j)^{1/2}$. The
relationship $-n_a U^a = (1+U_i h^{ij} U_j)^{1/2}$ is proved by inserting
the space--time split of the inverse metric $g^{ab}$ into the identity
$-1 = U_a g^{ab} U_b$ and solving for $(1+U_i h^{ij} U_j)$. To summarize,
$U_i$ as it appears in equations (3.8) and (3.9) should be viewed as
shorthand notation for the right--hand--side of equation (3.7). Then
equations (3.8) and (3.9) implicitly determine $(hJ)^i$ and the
Lagrangian number density $n$ as functions of the Eulerian number density
$\Pi$, the fields $\varphi$, $\theta$, $s$ $\bA$, $\aA$, and the spatial
metric $h_{ij}$.

The Hamiltonian form of the action is now obtained by substituting
expression (3.8) for $(hJ)^i$ into the action (3.5) and using equation
(3.7). This yields
$$S = \int dt\,d^3x \Bigl\{ \Pi\dot\varphi + (\Pi s)\dot\theta + (\Pi\bA)
                 \dot\aA   -N\Hfluid - N^i\Hifluid \Bigr\} \ ,\eqno(3.10)$$
where the fluid contributions to the Hamiltonian and momentum constraints are
$$\eqalignno{ \Hfluid &= \sqrt{h} \bigl[ \rho(1+U^iU_i) + pU^iU_i \bigr]
       \ ,&(3.11a)\cr
             \Hifluid &= -\mu\Pi U_i \ .&(3.11b)\cr}$$
The corresponding perfect fluid Hamiltonian is
$$H^{\sss\rm fluid} := \int d^3x \bigl( N\Hfluid + N^i\Hifluid \bigr)
          \ .\eqno(3.12)$$
In equations (3.10--12), $n$ and $U_i$ are considered to be functions of
the variables $\Pi$, $\varphi$, $(\Pi s)$, $\theta$, $(\Pi\bA)$, $\aA$, and
$h_{ij}$ as determined by equations (3.7) and (3.9).

The fundamental Poisson brackets among the fluid variables are
$$\eqalignno{ \{\varphi(x),\Pi(x')\}   &= \delta(x,x') \ ,&(3.13a)\cr
              \{\theta(x),(\Pi s)(x')\}  &= \delta(x,x') \ ,&(3.13b)\cr
        \{\aA(x),(\Pi {\beta_{\sss B}}) (x')\} &=
        \delta^{\sss A}_{\sss B}\delta(x,x') \ .&(3.13c)\cr}$$
Observe that the fluid contributions (3.11) to the
Hamiltonian and momentum constraints are given by the projections
$\Hfluid=\sqrt{h} n_aT^{ab}n_b$ and $\Hifluid = \sqrt{h}n_a{T^a}_i$ of the
stress tensor (2.1), as expected of
a matter action with nonderivative coupling to gravity [17].
Correspondingly, the gravitational field contributions to the constraints
are the same as for vacuum general relativity. The fluid contribution
(3.11a) to the Hamiltonian constraint can be rewritten in various useful
forms by using expression (3.9) to relate the Eulerian and Lagrangian
number densities:
$$\eqalignno{\Hfluid &= \bigl[ (\mu\Pi)^2 + \Hifluid h^{ij} {{\cal
         H}_j^{\sss\rm fluid}}   \bigr]^{1/2} - \sqrt{h}p  &(3.14a)\cr
                 &= \mu \Pi^2 /(\sqrt{h} n) - \sqrt{h}p \ .&(3.14b)\cr}$$
Also observe that the momentum constraint (3.11b) equals
$$ \Hifluid = \Pi\varphi_{,i} + (\Pi s)\theta_{,i} + (\Pi\bA){\aA}_{,i}
             \ ,\eqno(3.15)$$
which is the form dictated by the role of $\Hifluid$ as the generator of
spatial diffeomorphisms for the scalar fields $\varphi$, $\theta$, and $\aA$
and their conjugates.

%%%%%%%%%*%%%%%%%%%*%%%%%%%%%*%%%%%%%%%*%%%%%%%%%*%%%%%%%%%*%%%%%%%%%*%%%%%
\bigskip
\noindent{\it 3.2 Canonical equations of motion and symmetries}
\medskip
In order to compute the canonical fluid equations of motion, first vary
expression (3.14b) for $\Hfluid$ with respect to $n$, $s$, $\Pi$, and
$h_{ij}$. This yields
$$\eqalignno{ \delta\Hfluid =& \sqrt{h}\Bigl[ -\mu(1+U\cdot U) + n(U\cdot U)
          {\partial^2\rho\over\partial n^2} \Bigr] \delta n
           +\sqrt{h}\Bigl[ n(U\cdot U) {\partial^2\rho\over\partial
            n\partial s} + {\partial\rho\over\partial s} \Bigr] \delta s \cr
       &+ {2\over\sqrt{h}}{\mu\over n} \Pi\delta\Pi
         + {1\over2}\sqrt{h} \bigl[ -n\mu (1+U\cdot U) -p\bigr] h^{ij}
        \delta h_{ij}  \ ,&(3.16)\cr}$$
where $U\cdot U := U^iU_i$ and $U^i := h^{ij}U_j$, and expression (3.9)
is used. By varying equation (3.9) for $\Pi$, $\delta n$ is given in terms
of variations in the canonical variables by
$$\eqalignno{ 0 =& \sqrt{h}\Bigl[ -\mu(1+U\cdot U) + n(U\cdot U)
           {\partial^2\rho\over \partial n^2} \Bigr] \delta n
     - {1\over2} \sqrt{h} n\mu \bigl[ (1+U\cdot U) h^{ij} - U^iU^j\bigr]
             \delta h_{ij} \cr
    &+\sqrt{h} n(U\cdot U) {\partial^2\rho\over\partial n\partial s}\delta s
     + \mu (1+U\cdot U)^{1/2} \delta\Pi
     - \sqrt{h} nU^i\delta(\mu U_i) \ .&(3.17)\cr}$$
Combining these results gives
$$\eqalignno{ \delta\Hfluid = &-{1\over2} \sqrt{h} \bigl[ ph^{ij} + (\rho + p)
           U^i U^j \bigr]   \delta h_{ij} \cr
   &+ (1+U\cdot U)^{-1/2} \bigl[ f\delta\Pi + T\delta(\Pi s) - U^i
       \delta\Hifluid \bigr]  \ ,&(3.18)\cr}$$
where $U_i$ and $n$ are functions of the canonical variables as determined
by equations (3.7) and (3.9). In addition, the pressure $p$, temperature $T$,
and chemical free energy $f$ are functions of the canonical variables defined
through the relations $p = n(\partial\rho/\partial n) -\rho$, $T =
(\partial\rho/\partial s)/n$, and $f = (\partial\rho/\partial n) -Ts$.

The functional derivatives of the fluid Hamiltonian (3.12) follow from the
variation (3.18) and the variation of $\Hifluid$ from equation (3.15). In
canonical form the perfect fluid equations of motion  are
$$\eqalignno{\dot\varphi &= {\delta H^{\sss\rm fluid}\over\delta\Pi\hfill}
            = N(f-U^i\varphi_{,i})(1+U\cdot U)^{-1/2} + N^i\varphi_{,i}
           \ ,&(3.19a)\cr
  \dot\theta &= {\delta H^{\sss\rm fluid}\over\delta(\Pi s)\hfill}
    = N(T-U^i\theta_{,i})(1+U\cdot U)^{-1/2} + N^i\theta_{,i} \ ,&(3.19b)\cr
  \dot\aA &= {\delta H^{\sss\rm fluid}\over\delta(\Pi\bA)\hfill}
     = N(-U^i{\aA}_{,i})(1+U\cdot U)^{-1/2} + N^i{\aA}_{,i} \ ,&(3.19c)\cr
   \dot\Pi &= -{\delta H^{\sss\rm fluid}\over\delta\varphi\hfill }
     = - (N\sqrt{h} n U^i)_{,i} + (N^i\Pi)_{,i} \ ,&(3.19d)\cr
 (\Pi s)^{\textstyle\cdot} &= -{\delta H^{\sss\rm
fluid}\over\delta\theta\hfill}
     = - (N\sqrt{h} ns U^i)_{,i} + (N^i\Pi s)_{,i} \ ,&(3.19e)\cr
 (\Pi\bA)^{\textstyle\cdot} &= -{\delta H^{\sss\rm fluid}\over\delta\aA \hfill}
     = - (N\sqrt{h} n\bA U^i)_{,i} + (N^i\Pi\bA)_{,i}  \ .&(3.19f)\cr }$$
These are precisely equations (2.10), (2.9), (2.8f), (2.8b), (2.8c), and
(2.8e),
respectively, written in terms of the 3+1 decomposition (3.1), (3.8) for $J^a$.
Also observe that the variation (3.18) yields
$$ {\delta H^{\sss\rm fluid}\over\delta h_{ij}\hfill }
         = -{1\over2} N\sqrt{h} \bigl[p h^{ij} + (\rho + p) U^iU^j\bigr]
             \ ,\eqno(3.20)$$
which is the fluid contribution to the canonical equation of motion for
the spatial metric $h_{ij}$.

The global symmetries (2.18) described in section 2.2 appear in the canonical
formalism as transformations on the canonical variables generated through the
Poisson brackets by
$$Q[F] = \int d^3x \, \Pi F(\alpha,\beta,s) \ .\eqno(3.21)$$
This phase space functional is obtained from the charge (2.24) by using
definition (3.2) for $\Pi$. Note that the combination $\mu U_i$ from
equation (3.7) is invariant under these transformations, so from equation
(3.9) $n$ is invariant as well. It follows that $Q[F]$ has vanishing
Poisson brackets with the Hamiltonian and momentum constraints (3.11),
confirming that $Q[F]$ indeed generates a symmetry of the theory. Also
observe that the symmetry generators (3.21) close under the Poisson
brackets according to
$$ \{ Q[F_1], Q[F_2] \} = Q[F] \ ,\eqno(3.22)$$
where $F = (\partial F_1/\partial\aA)(\partial F_2/\partial\bA) -
(\partial F_1/\partial\bA)(\partial F_2/\partial\aA)$.

%%%%%%%%%*%%%%%%%%%*%%%%%%%%%*%%%%%%%%%*%%%%%%%%%*%%%%%%%%%*%%%%%%%%%*%%%%%
\bigskip
\noindent{\it 3.3 Initial and boundary value problems}
\medskip
A perfect fluid with equation of state $\rho(n,s)$ coupled to the
gravitational field is described by the canonical action (3.10) plus the
canonical action for gravity. The Cauchy data for this system consist of
the fluid variables $\varphi$, $\Pi$, $\theta$, $(\Pi s)$, $\aA$, $(\Pi\bA)$,
and the canonical gravitational variables.  These initial data cannot be
specified independently, but must satisfy the Hamiltonian and momentum
constraints. Any set of initial data that does satisfy these constraints
can be transformed into an equivalent set by the symmetries (2.18), which
are generated by the phase space functional $Q[F]$ of equation (3.21).

According to the analysis of section 2.2, the fields
$\varphi$ and $\theta$ can be brought to zero on any spacelike hypersurface
by a symmetry transformation. Thus, there is no loss of generality in
choosing $\varphi$ and $\theta$ to be zero on the initial hypersurface.
Moreover, the Lagrangian coordinates $\aA$ can be chosen to coincide with
the coordinates $x^i$ on the initial surface, so that ${\aA}_{,i} =
\delta_i^{\sss A}$. With these choices, equation (3.7) shows that the spatial
components $U_i$ of the covariant fluid four--velocity on the initial
hypersurface are
$$\mu U_i = -\beta_i \ .\eqno(3.23)$$
This reveals the geometrical significance of the fields $\bA$: with the
choices $\varphi = \theta=0$, ${\aA}_{,i} = \delta_i^{\sss A}$ allowed by
symmetry, the fluid space covector components $\bA$ are just $-\mu$ times
the spatial components of the fluid four--velocity.

Recall the definition $V_a := \mu U_a $ for the Taub current vector from
equation (2.15). The result (3.23) shows that in specifying initial data,
$-\bA$ can be identified with the spatial components $V_i$ of the Taub vector.
Thus, a complete set of initial data for a perfect fluid consists of the
(Eulerian) particle number density, the (Eulerian) entropy density, and the
spatial part of the Taub vector, along with $\varphi = \theta = 0$ and
$\aA = {\hbox{(spatial coordinates)}}$. These initial data are then evolved
according to the Hamiltonian differential equations of motion (3.19).

Now consider the boundary value problem. Assume the spacetime manifold
admits closed spacelike hypersurfaces so the boundary data is specified only
on initial and final hypersurfaces. One possible set of boundary data
consists in specifying the canonical coordinates $\varphi$, $\theta$, and
$\aA$ on the initial and final hypersurfaces. These boundary data include
$10\times\infty^3$ boundary values, $5\times\infty^3$ on the initial surface
and $5\times\infty^3$ on the final surface, where $\infty^3$ is the number
of space points. With these boundary data, the $10\times\infty^3$
Hamiltonian first order differential equations of motion (3.19) generically
determine a solution for the ten canonical field variables.

Another possible set of boundary data consists in specifying the canonical
momentum $\Pi$ along with the coordinates $\theta$ and $\aA$ on the initial
and final hypersurfaces. In this case, the data on the initial and final
surfaces are related through the conserved charges $Q[F]$ of equation
(3.21). In particular, if $\aA$ and $\Pi$ are specified initially and $\aA$
is specified finally, then the conserved charges $Q[F]$ can be used to
compute $\Pi$ on the final surface by considering a complete set of
functions $F(\aA)$. This means that the independent boundary data consist
of only $9\times\infty^3$ boundary values. With these boundary data, the
$10\times\infty^3$ Hamiltonian first order differential equations of
motion generically determine the ten canonical field variables to within
the symmetry transformation (2.19). In particular, the field $\varphi$
is obtained only to within an additive constant along each of the
$\infty^3$ flow lines. Other sets of boundary data are restricted by the
conserved charges $Q[F]$ as well.

The distinctions among the various types of boundary data
can be clarified by a simple example, namely, the free nonrelativistic
particle. If the initial and final positions of the particle are given as
boundary data, the equations of motion can be solved uniquely for the
particle position as a function of time. But the initial and final momenta
cannot be specified independently, because space translation invariance
implies that the momentum is conserved. By specifying equal values for the
intial and final momenta, the equations of motion can be solved to within a
constant spatial translation of the particle.

%%%%%%%%%*%%%%%%%%%*%%%%%%%%%*%%%%%%%%%*%%%%%%%%%*%%%%%%%%%*%%%%%%%%%*%%%%%
\bigskip
\noindent{\bf 4. Velocity--potential representation}
\medskip
In equation (2.8a), namely
$$U_a = -(\varphi_{,a} + s\theta_{,a} + \bA{\aA}_{,a})/\mu \ ,\eqno(4.1)$$
the fluid four--velocity is written in terms of various scalar fields and
their gradients. Expressions of this type are common in the literature on
fluid dynamics. They are often called velocity--potential representations
or Clebsch [29] representations of $U_a$, and the scalar fields themselves
are called velocity potentials or Clebsch potentials. Two related
questions naturally arise: Is the velocity--potential representation (4.1)
sufficiently general to allow for {\it any\/} four--velocity, or does it
restrict the four--velocity in some way?  If the representation (4.1) is
sufficiently general, then is it overly general in the sense that fewer
potentials would be adequate?

The first of these questions can be answered in the affirmative by
explicitly constructing a set of velocity potentials $\varphi$, $\theta$,
$\aA$, and $\bA$ that correspond to an arbitrary timelike, unit
normalized vector field $U^a$, along with arbitrary spacetime scalar
fields $s$, $T$, and $\mu\neq 0$. Begin by defining the thermasy through
the relationship $T = \theta_{,a}U^a$. More precisely, assign arbitrary
values for $\theta$ on some spacelike hypersurface, then define the value
of $\theta$ at any other spacetime point $x$ by integrating the scalar $T$
along a flow line of $U^a$ from the hypersurface to the point $x$.
Likewise, choose arbitrary values for $\varphi$ on some spacelike
hypersurface then extend $\varphi$ off this hypersurface by integrating
the relationship $f:=\mu -Ts = \varphi_{,a}U^a$ along the flow lines of
$U^a$. Now observe that the vector
$$\beta_a := -(\mu U_a + \varphi_{,a} + s\theta_{,a}) \eqno(4.2)$$
is orthogonal to the flow lines of $U^a$; that is, $\beta_a U^a =0$. Thus,
$\beta_a$ can be expressed as a linear combination of three independent
basis vector fields, where the basis vectors span the subspace of
spacetime vector fields that are othogonal to $U^a$. Such basis vectors
are the gradients of three scalar fields $\aA$, where $\aA$ are a set
of Lagrangian coordinates that (uniquely) label the flow lines of $U^a$.
Therefore $\beta_a$ can be written as $\beta_a = \bA{\aA}_{,a}$ where
$\bA$ are scalar fields. Inserting this expression for $\beta_a$ into
equation (4.2) yields the velocity--potential representation (4.1).

The construction described above shows that the fluid four--velocity
$U_a$ always can be expressed in the velocity--potential representation
(4.1) for any entropy per particle $s$, any nonzero chemical potential
$\mu$, and any temperature $T$ such that the thermasy satisfies
$T=\theta_{,a}U^a$. This construction also shows the extent to which
the velocity potentials are arbitrary. In particular, $\varphi$ and
$\theta$ can be chosen arbitrarily on a single spacelike hypersurface,
and the fields $\aA$ can be chosen as any set of unique flow line
labels. These ambiguities in the velocity--potential representation
appear as the invariances (2.19--21) of the perfect fluid action.

Regarding the second question, it is indeed possible to reduce the number of
potentials that appear in the velocity--potential representation by
replacing the three pairs of fields $\aA$, $\bA$, by a single pair (see
references [4, 5, 15]). The demonstration invokes Pfaff's theorem [37] and
the observation from section 2.2 that $\bA$ can be viewed as a covector or
one--form on the three--dimensional fluid space. According to Pfaff's
theorem, $\bA$ can be written in terms of three fluid space scalar fields
as $\bA = \tilde\varphi_{,{\sss A}} + \bar\beta {\bar\alpha}_{,{\sss A}} $.
Then the velocity--potential representation (4.1) becomes
$$U_a = -({\bar\varphi}_{,a} + s\theta_{,a} +\bar\beta{\bar\alpha}_{,a})/\mu
          \ ,\eqno(4.3)$$
where the definition $\bar\varphi := \varphi + \tilde\varphi$ has been used.
Comparing equations (4.1) and (4.3) shows that the seven potentials
$\varphi$, $\aA$, $\bA$, have been replaced by three potentials
$\bar\varphi$, $\bar\alpha$, $\bar\beta$. In effect, this reduction amounts
to restricting the fluid space indices on $\aA$ and $\bA$ to a single value.

The perfect fluid action (2.5) admits the symmetry transformations (2.18)
regardless of the range of values assumed by the fluid space indices.
However, the identification of $\aA$ as Lagrangian coordinates was used
repeatedly in the interpretation of those symmetries and the corresponding
conserved charges (2.24). It should be emphasized that the reduction in
the number of velocity potentials precludes this possibility of identifying
$\aA$ as Lagrangian coordinates for the fluid. Thus, with just one
pair of variables in place of $\aA$ and $\bA$, the physical interpretation
of the symmetries and conserved charges is lost. Also recall that with
three pairs $\aA$, $\bA$, the fields $\bA$ can be given a direct
geometrical interpretation by relating them to the spatial components
of $-V_a$ to within a symmetry transformation (see section 3.3). This
allows for the specification of initial data in terms of simple physical
quantities, namely the Eulerian number density, the Eulerian entropy
density, and the spatial part of the Taub current.

Another disadvantage of the reduced representation (4.3) is that the velocity
potentials $\bar\varphi$, $\bar\alpha$, $\bar\beta$ are not always single
valued functions on  spacelike hypersurfaces. (This has been recognized in
the context of nonrelativistic perfect fluids in reference [38].) For
example, consider a fluid whose four--velocity is described by the
representation (4.1) with $\beta_1 = -\alpha^2$, $\beta_2 = \alpha^1$,
$\beta_3 = 0$. Such a fluid has nonzero vorticity with axis of rotation
(in the isentropic case) in the direction $\epsilon^{abcd}\varphi_{,b}
{\alpha^1}_{,c}{\alpha^2}_{,d}$. Using the Pfaff reduction, $\bA$ can be
written as $\bA = \tilde\varphi_{,{\sss A}} + \bar\beta
{\bar\alpha}_{,{\sss A}} $ where $\tilde\varphi = R^2\phi$, $\bar\alpha
= R^2$, and $\bar\beta = -\phi$. Here, $R$ and $\phi$ are polar
coordinates in the $\alpha^1$--$\alpha^2$ plane of the fluid space,
defined by $R^2 = (\alpha^1)^2 + (\alpha^2)^2$ and $\tan\phi =
\alpha^2/\alpha^1$. Since the value of $\phi$ jumps by $2\pi$ along a
spacelike loop surrounding the vorticity axis, the velocity potentials
$\tilde\varphi$ (and hence $\bar\varphi$) and $\bar\beta$ are not single
valued functions on space.

This example shows that the representation (4.3), with potentials
$\bar\varphi$, $\bar\alpha$, $\bar\beta$ that are single valued, does
{\it not\/} allow for an arbitrary four--velocity. As a particular
consequence, it can be shown [38, 39] that the fluid helicity is
restricted to vanish. (Helicity is defined, for example, in references
[34] and [40].) On the other hand, the velocity--potential
representation (4.1) can be used to represent any four--velocity with
potentials that are single valued on all spacelike hypersurfaces. Since
several coordinate charts may be needed to cover the fluid space with
Lagrangian coordinates $\aA$, a more precise statement is that any
four--velocity can be represented as in equation (4.1) by velocity
potentials that are single valued within open subsets of spacelike
hypersurfaces, where the open subsets contain flow lines $\aA$ from a
single fluid space coordinate chart.

In order to verify this claim, recall the discussion at the beginning of
this section in which an arbitrary four--velocity $U_a$ was constructed
from the velocity--potential representation (4.1).  The goal now is to
argue that the potentials used in that construction are single valued.
Assuming $\varphi$ and $\theta$ are chosen to be single valued on one
spacelike hypersurface, then integration of the equations
$f=\varphi_{,a}U^a$ and $T=\theta_{,a}U^a$ yields spacetime fields
$\varphi$ and $\theta$ that are single valued on any spacelike
hypersurface. Of course, $s$ is always single valued since it has a
direct physical interpretation. The Lagrangian coordinates $\aA$ from a
given coordinate chart on the fluid space are also single valued
functions on any spacelike hypersurface. From equation (4.2), the fields
$\bA$ on a hypersurface with coordinates $x^i$ are now defined by
$$\bA = - \alpha_{{\sss A}}^i(\mu U_{i} + \varphi_{,i} + s\theta_{,i})
             \ ,\eqno(4.4)$$
where $\alpha_{{\sss A}}^i$ is the inverse of ${\aA}_{,i}$. It follows
that $\bA$ are single valued functions on any hypersurface since each
field appearing on the right--hand--side of equation (4.4), including the
spatial components $U_i$ of the fluid four--velocity, are single valued.
The conclusion is that within a subset of any spacelike hypersurface that
contains flow lines from a given fluid space coordinate chart, the
velocity--potentials appearing in the representation (4.1)
can be assumed to be single valued without any restriction on $U_a$.

Another consequence of reducing the number of velocity potentials is that
the counting of degrees of freedom (per space point) is changed. With all
three pairs $\aA$, $\bA$, the perfect fluid is described by five pairs of
unconstrained canonical variables and therefore has the expected five degrees
of freedom. Three of these correspond to the fluid's freedom of motion in
three--dimensional space, while the remaining two degrees of freedom
correspond to the fluid number density and entropy density. With just one
pair of potentials replacing $\aA$ and $\bA$, the system apparently has just
three degrees of freedom. Evidently, this difference in the number of degrees
of freedom arises because in the steps leading to the reduced representation
(6.3), the equations of motion ${\bA}_{,a}U^a = 0$ and ${\aA}_{,a}U^a = 0$
were used in order to interpret $\bA$ as a fluid space covector. The
significance of using the equations of motion can be understood by an
analogy: Consider a nonrelativistic particle in three spatial dimensions
moving in a central potential. The particle has three degrees of freedom,
corresponding to motion in each of the three spatial directions. But
the equations of motion show that the particle actually moves in a
two--dimensional plane. Using this result, the problem can be reduced to
that of a particle moving in its orbital plane, and the number of degrees
of freedom is reduced to two.

Finally, observe that if the term $\varphi_{,a}$ had been omitted from
the representation (4.1), the Pfaff reduction would nevertheless lead to
the representation (4.3) with an apparently trivial change of notation
$\bar\varphi\to\tilde\varphi$. This reasoning  has incorrectly lead to
the conclusion (see the appendix of reference [5]) that the potential
$\varphi$ is unnecessary, and that a term $\varphi_{,a}J^a$ is not needed
for a valid action principle if all three pairs of fields $\aA$, $\bA$ are
used. The conclusion is not valid because $\tilde\varphi$ is a function on
the fluid space, and satisfies $\tilde\varphi_{,a}U^a =0$. This
contradicts the result obtained by contracting equation (4.3) (with
$\bar\varphi\to\tilde\varphi$) with $U^a$, which shows that
$\tilde\varphi_{,a}U^a$ equals the chemical free energy $f$.
Another (invalid) justification [5] for dropping $\varphi$
is that the variation of $\varphi$ in the action just yields ${J^a}_{,a}=0$,
and particle number conservation is already implied by the constancy of
the labels $\aA$ along the fluid flow lines. The error in logic appears
to lie in the interpretation of $\aA$ as {\it particle\/} labels,
in which case their existence would imply conservation of particle number.
However, true particle labels are not continuous spacetime fields, so
$\aA$ do not label individual particles. Rather, $\aA$ label the fluid
flow lines that are physically determined by the average particle motion.
The particle number conservation equation ${J^a}_{,a}=0$ must be imposed
as a separate equation of motion to insure that the number of particles
within a flow tube (defined by a bundle of flow lines) is conserved.

%%%%%%%%%*%%%%%%%%%*%%%%%%%%%*%%%%%%%%%*%%%%%%%%%*%%%%%%%%%*%%%%%%%%%*%%%%%
\bigskip
\noindent{\bf 5. Action $\bar S$ with equation of state $\rho(n,s)$}
\medskip
\noindent{\it 5.1 Action and equations of motion}
\medskip
Express the densitized fluid number flux vector as
$$J^a = -\sqrt{-g} \epsilon^{abcd}\eta_{123}(\alpha) \,{\alpha^1}_{,b}
     {\alpha^2}_{,c} {\alpha^3}_{,d} \ ,\eqno(5.1)$$
where $\eta_{123}$ is a function of $\alpha$. The significance of
$\eta_{123}(\alpha)$ can be seen by using the number flux vector
$nU^a = J^a/\sqrt{-g}$ at a spacetime point $x$ to construct a differential
three--form on the spacelike hypersurface orthogonal to the fluid flow line
at $x$. This three--form is
$$ {1\over 3!} n U^a\epsilon_{abcd} dx^b\wedge dx^c\wedge dx^d =
    \eta_{123} {d\alpha}^1\wedge {d\alpha}^2\wedge {d\alpha}^3
      \ ,\eqno(5.2)$$
and is interpreted as the number of particles in the infinitesimal
three--volume $(1/3!)U^a\epsilon_{abcd} dx^b\wedge dx^c\wedge dx^d$. Then
equation (5.2) shows that $\eta_{123}(\alpha)$ is the component of a
three--form $\eta = (1/3!)\eta_{{\sss ABC}} {d\alpha}^{{\sss A}}\wedge
{d\alpha}^{{\sss B}}\wedge {d\alpha}^{{\sss C}}$ on the fluid space
whose integral over a region $B$ gives the number of fluid particles whose
flow lines are included in $B$:
$$\int_B \eta = ({\hbox{number of particles in $B$}}) \ .\eqno(5.3)$$
Under changes of coordinates (2.21a) in the fluid space, $\eta_{123}$
transforms as the component of a three--form. Correspondingly, expression
(5.1) for $J^a$ is independent of the choice of fluid space coordinates.
Note that {\it locally\/}, $\eta_{123}$ can be set to unity by an
appropriate choice of coordinates.

The equations of motion (2.8b,f), expressing conservation of particle
number and constancy of the Lagrangian coordinates $\aA$ along the fluid
flow lines, are immediately satisfied by virtue of the ansatz (5.1) for
$J^a$. If the entropy per particle $s$ is given as a function of $\aA$, the
entropy exchange constraint (2.8c) is also automatically satisfied. Observe
that $s\eta$ is the fluid--space three--form whose integral over a region
$B$ gives the total entropy contained in the flow lines included in $B$.

Given a three--form $\eta$, a function $s$, and coordinates $\aA$ on the
fluid space, a perfect fluid action $\bar S$ can be constructed that is a
functional of the Lagrangian coordinates $\aA$ only. First consider the
action $S' = S - \int d^4x (\varphi J^a + s\theta J^a)_{,a}$, that differs
from $S$ of equation (2.5) by boundary terms. This action $S'$ yields the
equations of motion (2.8) that correctly describe a relativistic perfect
fluid with equation of state $\rho(n,s)$. From the action $S'$, construct
the functional $\bar S$ by substituting expression (5.1) for $J^a$ and the
function $s(\alpha)$ for $s$. The terms $-\varphi {J^a}_{,a} - \theta
(sJ^a)_{,a} + \bA{\aA}_{,a}J^a$ drop out, leaving the result
$$\bar S = -\int d^4x \, \sqrt{-g} \rho(|J|/\sqrt{-g}, s) \ ,\eqno(5.4)$$
where $s=s(\alpha)$ and $J^a$ is defined as a function of $\aA$ through
equation (5.1). The procedure used to obtain this functional (5.4) does not
guarantee its validity as an action principle. In particular, the equations
of motion that follow by varying $\bar S$ with respect to $\aA$ are not
necessarily equivalent to the original equations of motion from $S$ or $S'$
with $J^a$ expressed as a function of $\aA$ through equation (5.1). Such a
discrepancy would occur if the variations in $J^a$ induced by variations in
$\aA$ are not the most general possible variations consistent with the
constraint equations ${J^a}_{,a}=0$, $(sJ^a)_{,a}=0$, and ${\aA}_{,a}J^a=0$.
However, in this case the representation (5.1) for $J^a$ is general, and the
functional (5.4) does constitute a valid perfect fluid action. This can be
confirmed explicitly by varying the functional (5.4) and showing that the
correct equations of motion follow. (Alternatively, the validity of the
functional (5.4) as an action would be assured if it could be demonstrated
that expression (5.1) and $s=s(\alpha)$ arise from solving the equations of
motion (2.8a--d,f) for $J^a$, $\varphi$, $\theta$, $s$ and $\bA$ in terms of
$\aA$.)

In varying the functional (5.4), observe first that variations with respect
to the metric $g_{ab}$ are unchanged by the substitution (5.1) for $J^a$. So
$\bar S$ yields the correct stress--energy tensor (2.1). Variation with
respect to $\aA$ gives
$$\delta \bar S = \int d^4x [\mu U_a \delta J^a(\alpha) - \sqrt{-g}
    nT \delta s(\alpha) ] \ ,\eqno(5.5)$$
where
$$\eqalignno{\delta J^a(\alpha) &= -{1\over3!} \sqrt{-g} \epsilon^{abcd}
        \biggl[  3\eta_{\sss ACD} \delta{\alpha^{\sss A}}_{,b}
        {\alpha^{\sss C}}_{,c}  {\alpha^{\sss D}}_{,d} + \biggl(
       {\partial \eta_{\sss BCD}\over\partial\alpha^{\sss A}}\biggr)
      {\alpha^{\sss B}}_{,b} {\alpha^{\sss C}}_{,c} {\alpha^{\sss D}}_{,d}
        \delta\alpha^{\sss A} \biggr] \cr
    &=  -{1\over3!} \sqrt{-g} \epsilon^{abcd} \biggl[ 3 (\eta_{\sss ACD}
       \delta\alpha^{\sss A} )_{,b} {\alpha^{\sss C}}_{,c}
              {\alpha^{\sss D}}_{,d} \cr
    &\qquad\qquad\qquad\qquad + \biggl( {\partial\eta_{\sss BCD}
       \over\partial\alpha^{\sss A}} -3 {\partial\eta_{\sss ACD}
       \over\partial\alpha^{\sss B}} \biggr) {\alpha^{\sss B}}_{,b}
       {\alpha^{\sss C}}_{,c} {\alpha^{\sss D}}_{,d}
        \delta\alpha^{\sss A}  \biggr]\ .&(5.6)\cr}$$
The last term above is proportional to the four--form
$4\eta_{{\sss [BCD},{\sss A]}} = \eta_{{\sss BCD},{\sss A}} -
3\eta_{{\sss A[CD},{\sss B]}}$, and must vanish because the fluid space
is three--dimensional. Inserting $\delta J^a(\alpha)$ into the variation
(5.5) and integrating by parts yields the equation of motion
$$0 = {1\over\sqrt{-g}} {\delta\bar S\over\delta\alpha^{{\sss A}}}
    = {1\over2}\epsilon^{abcd} V_{a;b} \eta_{{\sss ACD}}
          {\alpha^{{\sss C}}}_{,c}
       {\alpha^{{\sss D}}}_{,d} - nTs_{,{\sss A}} \ ,\eqno(5.7)$$
where $V_a = \mu U_a$ is the Taub current. Equation (5.2) implies
$\eta_{{\sss ACD}} {\alpha^{\sss A}}_{,a} {\alpha^{\sss C}}_{,c}
{\alpha^{\sss D}}_{,d} = nU^b \epsilon_{bacd}$, so the equation of motion
simplifies to
$$\eqalignno{ 0 &= {1\over2} \epsilon^{efcd} V_{e;f} U^b\epsilon_{bacd} -
          Ts_{,a} \cr
            &= 2 V_{{\sss [}a;b{\sss ]}} U^b - T s_{,a} \ .&(5.8)\cr}$$
This is the combination (2.14) of the Euler equation and the entropy
conservation equation. This analysis shows that the functional $\bar S $
is indeed a valid perfect fluid action: $\bar S$ has the correct stress
tensor, incorporates particle number conservation and entropy conservation
by virtue of expressions (5.1) for $J^a$ and $s=s(\alpha)$, and yields
the Euler equation from its equations of motion.

%%%%%%%%%*%%%%%%%%%*%%%%%%%%%*%%%%%%%%%*%%%%%%%%%*%%%%%%%%%*%%%%%%%%%*%%%%%
\bigskip
\noindent{\it 5.2 Hamiltonian form of the action $\bar S$}
\medskip
The momenta conjugate to $\aA$ are
$$\eqalignno{ P_{\sss A} &= - {1\over2} \sqrt{-g} \mu U_a\epsilon^{atcd}
  \eta_{\sss ACD} {\alpha^{\sss C}}_{,c} {\alpha^{\sss D}}_{,d} &(5.9a)\cr
       &= -\mu\Pi U_i \alpha^i_{\sss A} \ .&(5.9b)\cr}$$
Here, $\alpha^i_{\sss A}$ is the matrix inverse of ${\alpha^{\sss A}}_{,i}$
and $\Pi := J^t$ is the spatially densitized Eulerian particle number
density (3.9):
$$\Pi = \sqrt{h} n (1+U_ih^{ij}U_j)^{1/2} \ .\eqno(5.10)$$
In principle, the
next step in deriving the Hamiltonian is to solve equations (5.9) for
$\dot\aA$ as functions of $P_{\sss A}$ and $\aA$. This is not possible for
arbitrary equations of state $\rho(n,s)$. The situation here is closely
analogous to the expression of the Hamiltonian for the action $S$,
discussed in section 3. In that case the Hamiltonian (3.11--12) is given
as a function of the Lagrangian number density $n$ and the spatial
components $U_i$ of the fluid velocity. In turn, $n$ and $U_i$ are
determined as functions of the canonical variables through equations (3.7)
and (3.9). For the case at hand, the Hamiltonian is again expressed by
equations (3.11) and (3.12), namely
$$ H^{\sss\rm fluid} = \int d^3x \Bigl\{ N\sqrt{h}\bigl( \rho(1+U^iU_i)
   + pU^iU_i\bigr)  + N^i\bigl(-\mu\Pi U_i\bigr) \Bigr\} \ .\eqno(5.11)$$
This result is dictated by the fact that the fluid contributions to the
Hamiltonian and momentum
constraints are just the energy and momentum densities of the fluid and are
given by appropriate projections of the fluid stress--energy--momentum tensor
[17]. The quantities $n$ and $U_i$ that appear in the Hamiltonian (5.11) are
determined as functions of the canonical variables $\aA$, $P_{\sss A}$
through equations (5.9b) and (5.10). For this purpose observe that,
according to expression (5.1),
$$\Pi = {1\over3!}\sqrt{h} \epsilon^{ijk} \eta_{\sss ABC}
     {\alpha^{\sss A}}_{,i}
    {\alpha^{\sss B}}_{,j}{\alpha^{\sss C}}_{,k} \ .\eqno(5.12)$$
Thus, $\Pi$  only depends on $\aA$ and their spatial derivatives, not on
$\dot\aA$. Also note that the fluid contribution to the momentum constraint
equals  $P_{\sss A}{\alpha^{\sss A}}_{,i}$ and is the canonical generator
of spatial diffeomorphisms for the scalar fields $\aA$ and their conjugates
$P_{\sss A}$.

The canonical equations of motion follow from the variation (3.18) of the
fluid contribution to the momentum constraint, where $\Pi$ is the function
(5.12) of $\aA$. This calculation shows that once again $\dot\aA$ is given
by equation (3.19c), and that $\dot P_{\sss A}$ is given by
$$\eqalignno{ \dot P_{\sss A} = -{\delta H^{\sss\rm fluid}\over
        \delta\aA\hfill}
   =& -\biggl({ NP_{\sss A}U^i\over(1+U\cdot U)^{1/2} }\biggr)_{,i}
        + (N^i P_{\sss A})_{,i} \cr
    &- N\sqrt{h}Tns_{,{\sss A}} + \Pi\alpha^i_{\sss A}
     \biggl({ N\mu\over(1+U\cdot U)^{1/2} }\biggr)_{,i} \ .&(5.13)\cr}$$
Equation (5.13) is just the space--time decomposition of the Lagrangian
equation of motion (5.7). Also note that the first two terms in $\dot
P_{\sss A}$ above coincide with $(\Pi\bA)^{\textstyle\cdot}$ from
equation (3.19f).

As mentioned in the introduction, the canonical Hamiltonian formulation of
perfect fluids derived here is related to the Lie--Poisson Hamiltonian
formulation by a Lagrangian to Eulerian map (see [18] and references
therein). The result of this mapping is a Hamiltonian description of
perfect fluids in which $\Pi$, $s$, and the fluid momentum density are the
fundamental variables. Since the Lie--Poisson brackets are not canonical,
there is no corresponding action functional of the form
``$\int(p\dot q - H)$".

%%%%%%%%%*%%%%%%%%%*%%%%%%%%%*%%%%%%%%%*%%%%%%%%%*%%%%%%%%%*%%%%%%%%%*%%%%%
\bigskip
\noindent{\it 5.3 Symmetries of the action $\bar S$}
\medskip
For fixed fluid space tensors $\eta$ and $s$, the local coordinate
expressions $s(\alpha)$ and
$\eta_{123}(\alpha)\,d\alpha^1\wedge d\alpha^2\wedge d\alpha^3$
can be constructed from any set of coordinates $\aA$
on the fluid space. Under a change of coordinates $\delta\aA =
\xi^{\sss A}(\alpha)$, the functions $\eta_{123}(\alpha)$ and $s(\alpha)$
at a given value of $\aA$ (at a given coordinate location) change
according to $-({\hbox{\it\$}}_\xi \eta)_{123}(\alpha)$ and
$-({\hbox{\it\$}}_\xi s)(\alpha)$, respectively, and correspondingly the
tensors $\eta_{123}(\alpha)\,d\alpha^1\wedge d\alpha^2\wedge d\alpha^3$
and $s(\alpha)$ at a given point on the fluid space manifold remain
unchanged. (Here, ${\hbox{\it\$}}_\xi$ is the Lie derivative in the
fluid space along the vector field $\xi$.) In this sense the functional
$\bar S$, which depends on $\aA$ only through the combinations
$\eta_{123}(\alpha)\,d\alpha^1\wedge d\alpha^2\wedge d\alpha^3$ and
$s(\alpha)$, is invariant under changes of fluid space coordinates.
Note however that this invariance of $\bar S$ involves a transformation
of the functions $\eta_{123}(\alpha)$ and $s(\alpha)$ at a given value
of $\aA$, as well as a transformation of the field variables $\aA$.
Consequently this invariance does not in general correspond to a
conserved Noether current. On the other hand, the subset of fluid space
coordinate transformations $\delta\aA = \xi^{\sss A}$ that satisfy
${\hbox{\it\$}}_\xi \eta =0$ and ${\hbox{\it\$}}_\xi s =0$ do not
involve a transformation of the functions $\eta_{123}(\alpha)$ and
$s(\alpha)$ at a given value of $\aA$ and do give rise to conserved
Noether currents. This same conclusion can be reached by considering
the changes in $\eta_{123}(\alpha)\,d\alpha^1\wedge d\alpha^2\wedge
d\alpha^3$ and $s(\alpha)$ induced by a general field transformation
$\delta\aA = \xi^{\sss A}$, where $\eta_{123}(\alpha)$ and $s(\alpha)$
are treated as fixed functions of $\aA$. Those changes are given by
$$\eqalignno{ \delta \bigl(\eta_{123}(\alpha)\,d\alpha^1\wedge
        d\alpha^2\wedge  d\alpha^3\bigr)
        &= ({\hbox{\it\$}}_\xi \eta)_{123}(\alpha) d\alpha^1\wedge
               d\alpha^2\wedge  d\alpha^3 \ ,&(5.14a)\cr
    \delta\bigl( s(\alpha)\bigr) &= ({\hbox{\it\$}}_\xi s)(\alpha)
      \ ,&(5.14b)\cr}$$
and show that $\bar S$ is invariant under transformations
$\delta\aA = \xi^{\sss A}$ that satisfy ${\hbox{\it\$}}_\xi \eta =0$
and ${\hbox{\it\$}}_\xi s =0$. The Noether current associated with
such transformations is obtained from a general variation of the
action, which is
$$\eqalignno{ \delta\bar S =& \ ({\hbox{terms giving the equations of
          motion}}) \cr
       &+ {1\over2}\int d^4x \bigl[ \sqrt{-g}\mu U_b \epsilon^{abcd}
      \eta_{\sss ACD} \delta\alpha^{\sss A} {\alpha^{\sss C}}_{,c}
       {\alpha^{\sss D}}_{,d}   \bigr]_{,a} \ .&(5.15)\cr}$$
For the transformations $\delta\aA = \xi^{\sss A}$ that leave
$\bar S$ invariant, the last integral in expression (5.15) must vanish
when the equations of motion hold and the conserved current is
$${1\over2} \sqrt{-g}\mu U_b  \epsilon^{abcd} \eta_{\sss ACD}
       \xi^{\sss A}(\alpha) {\alpha^{\sss C}}_{,c} {\alpha^{\sss D}}_{,d}
         \ .\eqno(5.16)$$
The Noether charge is the integral of the time component of this current
over a spacelike hypersurface. Using the definition (5.9) for the conjugate
momenta, this charge can be written as
$$Q[\xi] = \int_\Sigma d^3x\, P_{\sss A} \xi^{\sss A} \ .\eqno(5.17)$$
$Q[\xi]$ is independent of the spacelike hypersurface $\Sigma$ for any
fluid space vector $\xi$ that leaves the tensors $\eta$
and $s$ invariant under Lie transport.

%%%%%%%%%*%%%%%%%%%*%%%%%%%%%*%%%%%%%%%*%%%%%%%%%*%%%%%%%%%*%%%%%%%%%*%%%%%
\bigskip
\noindent{\bf 6. Other action functionals}
\medskip
\noindent{\it 6.1 Equation of state $p(\mu,s)$}
\medskip
A relativistic perfect fluid with equation of state $\rho(n,s)$ is described
by histories that extremize the action (2.5), along with the gravitational
action, under independent variations of $J^a$,
$\varphi$, $\theta$, $s$, $\aA$, $\bA$, and $g_{ab}$. This choice of
variables is convenient, but not unique. For example, replace $J^a/\sqrt{-g}$
by $nU^a$ with $n$ and $U^a$ varied separately, subject to the
normalization condition $U^aU_a = -1$. This changes the individual equations
of motion, but as a set the new equations are equivalent to the original
equations (2.8). In particular, the stress tensor has the standard perfect
fluid form (2.1) only when certain other equations of motion hold.

The action (2.5) with $J^a/\sqrt{-g}$ replaced by $nU^a$ is
$$S = -\int d^4x \sqrt{-g} \bigl\{ \rho(n,s) - nU^a (\varphi_{,a} -
     \theta s_{,a} +   \bA{\aA}_{,a} ) \bigr\} \ ,\eqno(6.1)$$
where $n$ is an independent variable. Now, according to the first law (1.3),
the chemical potential $\mu$ is the function of $n$ and $s$ defined by
$\mu = \partial\rho /\partial n$. If this equation could be inverted for $n$
as a function of $\mu$ and $s$, then $\mu$ could be used as an independent
variable in the action (6.1) by substituting $n(\mu,s)$ in place of $n$. In
effect, this inversion is accomplished by using the pressure $p$ as a
function of $\mu$ and $s$ to generate the functions $n(\mu,s)$, $\rho(\mu,s)$
through the first law in the form (1.4). That is, let $p(\mu,s)$ specify the
equation of state and determine $n$ and $\rho$ through
$$\eqalignno{ n &:= {\partial p\over\partial\mu} \ ,&(6.2a)\cr
              \rho &:= \mu {\partial p\over\partial\mu} - p \ .&(6.2b)\cr}$$
Inserting these relationships into the action (6.1) yields an action $S_{{\sss
(}p {\sss )}}$ for a perfect fluid with equation of state $p(\mu,s)$.

For the action $S_{{\sss (}p {\sss )}}$, it is convenient to use the Taub
vector $ V^a = \mu U^a$ as the independent variable in place of $\mu$ and
$U^a$. With this definition, the action with equation of state $p(\mu,s)$
reads
$$S_{{\sss (}p {\sss )}} = \int d^4x \sqrt{-g} \biggl\{  p - \Bigl({\partial
     p\over\partial \mu}\Bigr) \Bigl( |V| -    V^a(\varphi_{,a} +s
        \theta_{,a} + \bA{\aA}_{,a} )/|V| \Bigr) \biggr\} \ ,\eqno(6.3)$$
where $p=p(\mu,s)$ and $\mu := |V| = \sqrt{-V^a g_{ab} V^b}$.
The equations of motion obtained by varying $\varphi$, $\theta$, $s$, $\aA$,
and $\bA$ yield equations (2.8b--f) respectively. The equation of motion
obtained by varying $V^a$ is
$$\eqalignno{ 0 = {1\over\sqrt{-g}}{\delta S_{{\sss (}p {\sss )}} \over
     \delta V^a}  =& {1\over|V|^2} \Bigl({\partial^2 p\over\partial\mu^2}
        \Bigr) \Bigl( |V|^2 - V^b (\varphi_{,a} +s \theta_{,a} +
        \bA{\aA}_{,a}) \Bigr) V_a \cr
     &- {1\over|V|} \Bigl({\partial p\over\partial\mu}\Bigr)
       (\varphi_{,b} +s \theta_{,b}  + \bA{\aA}_{,b}) \Bigl(\delta^b_a
        + {V^bV_a\over|V|^2}\Bigr) \ .&(6.4)\cr}$$
The projection of this equation orthogonal to the fluid flow lines shows that
$(\varphi_{,a} +s \theta_{,a} + \bA{\aA}_{,a})$ is proportional to $V_a$, and
then  projection along the flow lines gives
$$V_a = -(\varphi_{,a} +s \theta_{,a} + \bA{\aA}_{,a}) \ .\eqno(6.5)$$
This is the equation of motion (2.8a), so the action $S_{{\sss (}p {\sss )}}$
indeed yields the complete set of fluid matter equations (2.8). As shown in
section 2, these equations of motion imply particle number conservation,
conservation of entropy along the flow lines, and the Euler equation. The
stress tensor derived from $S_{{\sss (}p {\sss )}}$ has the perfect fluid
form (2.1) when equation (6.5) holds.

The equation of motion (6.5) can be used to eliminate $V^a$ from the action
$S_{{\sss (}p {\sss )}}$ yielding an action functional
that is essentially the one found by Schutz [15]:
$$\tilde S_{{\sss (}p {\sss )}} = \int d^4x \sqrt{-g} p(\mu,s) \ .\eqno(6.6)$$
Here, $\mu$ is treated as the function of $\varphi$, $\theta$, $s$, $\aA$,
$\bA$, and $g_{ab}$ determined by $\mu^2 = {-V^aV_a}$, where $V_a$ is given by
equation (6.5). The only difference between the action (6.6) and Schutz's
action
is a difference in the number of pairs of fields $\aA$ and $\bA$. Schutz uses
just one pair, in accordance with the Pfaff reduction discussed in section 4.
The canonical form of Schutz's action is constructed in references [19]
and [20].

%%%%%%%%%*%%%%%%%%%*%%%%%%%%%*%%%%%%%%%*%%%%%%%%%*%%%%%%%%%*%%%%%%%%%*%%%%%
\bigskip
\noindent{\it 6.2 Equation of state $a(n,T)$}
\medskip
An action $S_{{\sss (}a {\sss )}}$ for a perfect fluid with equation of state
specified by $a(n,T)$, the physical free energy as a function of number
density and temperature, is obtained by eliminating the entropy per particle
$s$ from the action (2.5). Thus, consider solving the equation of motion
(2.8d) for $s$ as a function of $n$ and $T = \theta_{,a}U^a$, then
eliminating $s$ from the action (2.5) by substituting the result $s(n,T)$.
The term $-\sqrt{-g}\rho + s\theta_{,a}J^a$ becomes a function of
$n=|J|/\sqrt{-g}$ and $T=\theta_{,a}J^a/|J|$ that is identified with $na$,
where $a$ is the physical free energy (1.2b). The resulting action for a
perfect fluid with equation of state $a(n,T)$ is
$$ S_{{\sss (}a {\sss )}} = -\int d^4x \Bigl\{
     |J|\, a(|J|/\sqrt{-g}\,,\, \theta_{,a}J^a/|J|) - J^a(\varphi_{,a} +
       \bA{\aA}_{,a})  \Bigr\} \ .\eqno(6.7)$$
$S_{{\sss (}a {\sss )}}$ is a functional of $J^a$, $\varphi$,
$\theta$, $\aA$, $\bA$, and $g_{ab}$.

It is straightforward to show that the equation of motion $\delta
S_{{\sss (}a {\sss )}}/\delta J^a = 0$ is equivalent to equation (2.8a),
and the equations of motion obtained by varying $\varphi$, $\theta$, $\aA$,
and $\bA$ are just equations (2.8b,c,e,f). Of course, the equation of motion
(2.8d) is missing, since $S_{{\sss (}a {\sss )}}$ does not depend on $s$. But
with the identification of thermasy as in (2.9), this equation simply
reiterates the relationship dictated by the first law of thermodynamics and
is, in this sense, superfluous. Therefore the equations of motion derived
from the action $S_{{\sss (}a {\sss )}}$, along with the interpretation of
the variables in a manner that is consistent with the first law (1.5), are
complete in the sense that they imply particle number conservation,
conservation of entropy along the flow lines, and the Euler equation. In
addition, the stress tensor obtained from the action (6.7) has the perfect
fluid form (2.1).

The action discussed by Kijowski {\it et al.\/} [14] can be obtained from
the action $S_{{\sss (}a {\sss )}} - \int d^4x (\varphi J^a)_{,a}$ by
substituting expression (5.1) for $J^a$. The terms $-\varphi {J^a}_{,a} +
\bA{\aA}_{,a}J^a$ drop out, leaving
$$\bar S_{{\sss (}a {\sss )}} = -\int d^4x \Bigl\{
     |J|\, a(|J|/\sqrt{-g}\,,\, \theta_{,a}J^a/|J|) \Bigr\} \ .\eqno(6.8)$$
Variations with respect to the metric $g_{ab}$ and the thermasy $\theta$
are unchanged by this substitution, so $\bar S_{{\sss (}a {\sss )}}$ yields
the correct stress tensor (2.1) and the conservation of entropy (2.8c). A
calculation similar to the one appropriate for $\bar S$ shows that
variations of $\bar S_{{\sss (}a {\sss )}}$ with respect to the Lagrangian
coordinates $\aA$ yield equation (2.14), which implies the Euler equation.
The Hamiltonian form of $\bar S_{{\sss (}a {\sss )}}$ and the symmetries and
conserved charges can be found along the same lines as the analysis found
in section 5 for $\bar S$. The Hamiltonian form of this action also has been
considered in reference [41].

%%%%%%%%%*%%%%%%%%%*%%%%%%%%%*%%%%%%%%%*%%%%%%%%%*%%%%%%%%%*%%%%%%%%%*%%%%%
\bigskip
\noindent{\it 6.3 Isentropic fluids and dust}
\medskip
Isentropic fluids are perfect fluids with a constant entropy per particle $s$.
The first law of thermodynamics in the form (1.5) indicates that
isentropic fluids are described  by an equation of state of the form
$$a(n,T) = {\rho(n)\over n} - sT \ ,\eqno(6.9)$$
where $s$ is the constant value of the entropy per particle. Inserting this
equation of state into the action (6.7) yields a functional like $S$ of
equation (2.5), but with two differences: the function $\rho$ only depends
on $n=|J|/\sqrt{-g}$, and $s$ appears as a fixed constant, not a variable.
Thus, $s\theta$ can be absorbed into $\varphi$ by the change of variables
$\varphi' := \varphi + s\theta$, which reduces the combination $\varphi_{,a}
+ s\theta_{,a}$ that appears in the action to $\varphi'_{,a}$.
This leads to the isetropic fluid action
$$S_{\rm isentropic} = \int d^4x \Bigl\{ -\sqrt{-g}\, \rho(|J|/\sqrt{-g})  +
     \varphi'_{,a}J^a + \bA{\aA}_{,a}J^a  \Bigr\} \ ,\eqno(6.10)$$
which is a functional of $J^a$, $\varphi'$, $\aA$, $\bA$, and $g_{ab}$. In
terms of the new variable $\varphi'$, the equation of motion (2.8a) obtained
by varying $J^a$ becomes $0 = \mu U_a + \varphi'_{,a} + \bA{\aA}_{,a}$.
Contracting with $U^a$ gives $\mu = \varphi'_{,a}U^a$ and shows that
$\varphi'$ is a ``potential" for the chemical potential $\mu$.

Dust is a particular case of an isentropic fluid in which the energy
density $\rho$ is proportional to the number density $n$ and the
pressure (2.7) is zero. The proportionality constant is the rest
mass--energy of a fluid particle, which equals the chemical potential
$\mu = \rho/n$. Inserting the equation of state $\rho =
\mu n$ into the isentropic fluid action (6.10) yields
$$S_{\rm dust} = \int d^4x \Bigl\{
   -\mu |J|  + \varphi_{,a}J^a + \bA{\aA}_{,a}J^a  \Bigr\} \ ,\eqno(6.11)$$
where the prime has been dropped from $\varphi$. An alternative to this
dust action is obtained by replacing $|J|/\sqrt{-g}$ in the action
(6.11) with the function $[(|J|/\sqrt{-g})^2/n + n]/2$ and treating $n$ as
a new dynamical variable. This is justified because the action (6.11) is
recovered when $n$ is eliminated from the new action by using the solution
$n = |J|/\sqrt{-g}$ to the equation of motion for $n$. After introducing
the variable $n$ into the dust action,  $J^a$ can be eliminated by using
the $J^a$ equation of motion. This yields an action for dust that is a
functional of $n$, $\varphi$, $\aA$, and $\bA$, namely
$$S'_{\rm dust} = -{\mu\over2} \int d^4x\sqrt{-g} n (U_a g^{ab} U_b + 1)
     \ ,\eqno(6.12)$$
where $U_a := -(\varphi_{,a} + \bA{\aA}_{,a})/\mu$.

The Hamiltonian form of the dust action (6.11) or (6.12) is just the
canonical action (3.10) but with the terms containing $\theta$ and $s$
omitted:
$$S_{\rm dust} = \int dt\,d^3x \Bigl\{ \Pi\dot\varphi + (\Pi\bA)
             \dot\aA  -N{{\cal H}^{\sss\rm dust}} -
         N^i{{\cal H}_i^{\sss\rm dust}} \Bigr\} \ .\eqno(6.13)$$
The dust contributions to the Hamiltonian and momentum constraints can
be found explicitly, because the chemical potential appearing in equation
(3.7) for $U_i$ has no $n$ dependence. Thus, from equations (3.14a) and
(3.15),
$$\eqalignno{ {{\cal H}_i^{\sss\rm dust}} &= \Pi\varphi_{,i}
           + (\Pi\bA){\aA}_{,i} \ ,&(6.14a)\cr
    {{\cal H}^{\sss\rm dust}} &= \bigl[ (\mu\Pi)^2 +
    {{\cal H}_i^{\sss\rm dust}} h^{ij} {{\cal H}_j^{\sss\rm
          dust}}\bigr]^{1/2} \ .&(6.14b)\cr}$$
In reference [26], the action (6.12) and its canonical form (6.13) are
used to analyze the quantum theory of gravity coupled to dust.

The equation of state (6.9) also can be inserted into the action
functional (6.8). This yields an isentropic fluid action
$$\bar S_{\rm isentropic} = -\int d^4x \Bigl\{ \sqrt{-g}\, \rho(|J|/\sqrt{-g})
      \Bigr\}   \ ,\eqno(6.15)$$
where a total divergence (boundary term) has been discarded. Here, $J^a$ is a
function of the Lagrangian coordinates $\aA$ as given by equation (5.1).
Specialized to a pressureless dust equation of state, this action becomes
$\bar S_{\rm dust} = -\int d^4x \, \mu|J|$. In the canonical form of this
dust action, the dust contribution to the momentum constraint is
$P_{\sss A}{\aA}_{,i}$, and the dust contribution to the Hamiltonian
constraint is given by equation (6.14b) where
$\Pi$ is a function of $\aA$ through equation (5.12).

%%%%%%%%%%%%%%%%%%%%%%%%%%%%%%%%%%%%%%%%%%%%%%%%%%%%%%%%%%%%%%%%%%%%%%%
\bigskip
\noindent{\bf Acknowledgments}
\medskip
I would like to thank C.R. Evans, K.V. Kucha\v r, V. Moncrief, T. Piran,
and J.W. York for helpful discussions. I also thank the Center for Relativity
at The University of Texas at Austin, the Aspen Center for Physics, and
the Institute of Field Physics at the University of North Carolina for
hospitality during various stages of this work. Research support was
received from National Science Foundation grant number PHY--8908741 to the
University of North Carolina.

%%%%%%%%%%%%%%%%%%%%%%%%%%%%%%%%%%%%%%%%%%%%%%%%%%%%%%%%%%%%%%%%%%%%%%%
\bigskip
\noindent{\bf Appendix}
\medskip
The symmetries of the action (2.5) are transformations among the potentials
$\varphi$, $s$, $\theta$, $\bA$, and $\aA$ that leave the one--form
$-\mu U = d\varphi + s d\theta + \bA d\aA$ and the entropy per particle
$s$ invariant at each spacetime point. That is, the action satisfies
$S[\varphi, s, \theta, \bA, \aA] = S[\varphi', s', \theta', \bA', {\aA}']$
if the primed and unprimed variables are related by
$$\eqalignno{ d\varphi + s d\theta + \bA d\aA &= d\varphi' + s' d\theta' +
          \bA' d{\aA}' \ ,&(A.1a)\cr
        s &= s' \ .&(A.1b)\cr}$$
In general, the primed variables can be nonlocal functions of the unprimed
variables. Thus, for example, $\varphi'(x)$ can depend on the unprimed
variables at spacetime points other than $x$. The analysis that
follows is restricted to the case of ultralocal transformations in which
the primed variables at $x$ depend only on the unprimed variables at $x$.
In particular, the primed variables are not allowed to depend on the
derivatives of the unprimed variables.

Equation (A.1a) can be viewed as the expression of a canonical
transformation with generating function $\varphi - \varphi'$ for a
fictitious phase space with coordinates $\theta$, $\aA$ and momenta $s$,
$\bA$. In terms of the true phase space variables discussed in section 3,
the symmetry transformations (A.1) are given by
$$\eqalignno{ \Pi d\varphi + (\Pi s)d\theta + (\Pi\bA)d\aA &= \Pi' d\varphi'
           +   (\Pi s)'d\theta' + (\Pi\bA)'d{\aA}' \ ,&(A.2a)\cr
     \Pi &= \Pi' \ ,&(A.2b)\cr
     (\Pi s) &= (\Pi s)' \ .&(A.2c)\cr}$$
Equation (A.2a) expresses a Mathieu transformation among the canonical fluid
variables at each space point. (Mathieu transformations are canonical
transformations with zero generating function; that is, they preserve the
form $p_i dq^i$ [1].) The extra restrictions
(A.2b) and (A.2c) can be analyzed by computing Poisson brackets of the
primed variables with one another using the definition of the Poisson
brackets in terms of the unprimed variables. For example, condition
(A.2b) along with the Poisson bracket relationship $\{\varphi',\Pi'\} = 1$
lead to $\partial\varphi'/\partial\varphi = 1$; therefore,
$$\varphi' = \varphi + f \ ,\eqno(A.3)$$
where $f$ is independent of $\varphi$.
The Poisson bracket of $\varphi'$ and $(\Pi s)'$ along with condition (A.2c)
show that $f$ is independent of $\theta$. A similar analysis gives
$$\theta' = \theta + g \ ,\eqno(A.4)$$
where $g$, like $f$, is a function only of $\Pi$, $(\Pi s)$, $\aA$, and
$(\Pi\bA)$. It also follows from conditions (A.2b) and (A.2c) that
$\alpha'$ and $(\Pi\bA)'$ are independent of $\varphi$ and $\theta$.

Using the results (A.3) and (A.4), equation (A.2a) can be written as
$$ d(\Pi\F) = fd\Pi + gd(\Pi s) + (\Pi\bA) d\aA + {\aA}' d(\Pi\bA)'
     \ ,\eqno(A.5)$$
where $\F$ is defined by $\Pi\F = \Pi f + (\Pi s)g + (\Pi\bA)'{\aA}'$.
This relationship can be investigated by treating various combinations of
primed and unprimed variables as independent. For example, assume that
$(\Pi\bA)'$ as a function of $\Pi$, $(\Pi s)$, $\aA$, and $(\Pi\bA)$ can be
inverted for $(\Pi\bA)$ as a function of $\Pi$, $(\Pi s)$, $\aA$, and
$(\Pi\bA)'$. Then $\Pi$, $(\Pi s)$, $\aA$, and $(\Pi\bA)'$ can be chosen
as independent variables and equation (A.5) immediately yields
$$\eqalignno{\varphi' - \varphi &= f = \F + \Pi{\partial\F\over\partial
              \Pi}  \ ,&(A.6a)\cr
     \theta' - \theta &= g = \Pi {\partial\F\over\partial(\Pi s) }
               \ ,&(A.6b)\cr
     (\Pi\bA) &= \Pi {\partial\F\over\partial\aA} \ ,&(A.6c)\cr
     {\aA}' &= \Pi{\partial\F\over\partial(\Pi\bA)'} \ .&(A.6d)\cr}$$
These equations imply
$$ 0 = \Pi\bigl(\partial\F/\partial\Pi\bigr) + (\Pi s)
   \bigl(\partial\F/\partial(\Pi s) \bigr) + (\Pi\bA)
   \bigl(\partial\F/\partial(\Pi\bA)'\bigr) \ ,\eqno(A.7)$$
and show that $\F$ is homogeneous of degree zero in the momenta $\Pi$,
$(\Pi s)$, and $(\Pi\bA)'$. As a consequence, the functional dependence of
$\F$ can be expressed as $\F = \F(\aA, (\Pi s)/\Pi, (\Pi\bA)'/\Pi)$ and
equation (A.6a) can be written as
$$\varphi' = \varphi + \F - (\Pi s) {\partial\F\over\partial(\Pi s) }
     - (\Pi\bA)' {\partial\F\over\partial(\Pi\bA)'} \ .\eqno(A.8)$$
Other choices of independent variables lead to alternative sets of
transformations, in the same way as canonical transformations are
categorized into various types.

The infinitesimal symmetry transformations are obtained by setting
$$\F = \bigl(\aA (\Pi\bA)' \bigr)/\Pi + \epsilon F \ ,\eqno(A.9)$$
where $\epsilon$ is an infinitesimal parameter. Equations (A.8) and
(A.6b,c,d) then become
$$\eqalignno{ \delta \varphi &= \epsilon F - \epsilon (\Pi s)
        {\partial F\over\partial(\Pi s) } - \epsilon (\Pi\bA)
        {\partial F\over\partial(\Pi\bA)} \ ,&(A.9a)\cr
     \delta\theta &= \epsilon \Pi{\partial F\over\partial(\Pi s)}
          \ ,&(A.9b)\cr
     \delta(\Pi\bA) &= -\epsilon \Pi {\partial F\over\partial\aA}
          \ ,&(A.9c)\cr
     \delta\aA &= \epsilon \Pi {\partial F\over\partial(\Pi\bA)}
           \ ,&(A.9d)\cr }$$
where $F$ has functional dependence $F(\aA, (\Pi s)/\Pi, (\Pi\bA)/\Pi)$.
The factors of $\Pi$ can be eliminated from these relationships, leading
directly to the expression (2.18) of the infinitesimal symmetry
transformations in terms of the fluid potentials $\varphi$, $s$,
$\theta$, $\bA$, and $\aA$.
%%%%%%%%%%%%%%%%%%%%%%%%%%%%%%%%%%%%%%%%%%%%%%%%%%%%%%%%%%%%%%%%%%%%%%%%%%%%
\bigskip
\frenchspacing
\noindent{\bf References}
\medskip
\item{[1]} Lanczos C 1970 {\it The Variational Principles of Mechanics\/}
(Toronto: University of Toronto Press)
\item{[2]} Dyson F J 1989 {\it Am. J. Phys.\/} {\bf 58} 209; Hojman S A
and Shepley L C 1991 {\it J. Math. Phys.\/} {\bf 32} 142
\item{[3]} Misner C W, Thorne K S, and Wheeler J A 1973 {\it Gravitation\/}
(San Fransisco: Freeman)
\item{[4]} Seliger R L and Whitham G B 1968 {\it Proc. Roy. Soc.\/} A
{\bf 305} 1
\item{[5]} Schutz B F and Sorkin R 1977 {\it Ann. Phys.\/} {\bf 107} 1
\item{[6]} Taub A H 1954 {\it Phys. Rev.\/} {\bf 94} 1468
\item{[7]} Taub A H 1969 {\it Commun. Math. Phys.\/} {\bf 15} 235
\item{[8]} Hawking S W and Ellis G F R 1973 {\it The Large Scale
Structure of Space--Time\/} (Cambridge: Cambridge University Press)
\item{[9]} Lin C C 1963 Hydrodynamics of helium II {\it Liquid Helium\/}
ed G Careri (New York: Academic Press)
\item{[10]} Serrin J 1959 Mathematical principles of classical fluid
mechanics {\it Handbuch der Physik\/} vol 8 ed S Fl\"ugge and C Truesdell
(Berlin: Springer--Verlag)
\item{[11]} Carter B 1973 {\it Commun. Math. Phys.\/} {\bf 30} 261
\item{[12]} Eckart C 1960 {\it Phys. Fluids\/} {\bf 3} 421
\item{[13]} Schmid L A 1970 {\it Pure Appl. Chem.\/} {\bf 22} 493
\item{[14]} Kijowski J, Sm\'olski A, and G\'ornicka A 1990 {\it Phys.
Rev.\/} D {\bf 41} 1875; Kijowski J and Tulczyjew W M 1979 {\it A Symplectic
Framework for Field Theories\/} (Berlin: Springer--Verlag)
\item{[15]} Schutz B F 1970 {\it Phys. Rev.\/} D {\bf 2} 2762
\item{[16]} Bombelli L and Torrence R J 1990 {\it Class. Quantum Grav.\/}
{\bf 7} 1747
\item{[17]} Kucha\v r K V 1976 {\it J. Math. Phys.\/} {\bf 17} 801
\item{[18]} Bao D, Marsden J, and Walton R 1985 {\it Commun. Math. Phys.\/}
{\bf 99} 319; Holm D D 1989 Hamiltonian techniques for relativistic fluid
dynamics and stability theory {\it Relativistic Fluid Dynamics\/} ed
A Anile and Y Choquet--Bruhat (Berlin: Springer--Verlag)
\item{[19]} Schutz B F 1971 {\it Phys. Rev.\/} D {\bf 4} 3559
\item{[20]} Demaret J and Moncrief V 1980 {\it Phys. Rev.\/} D {\bf 21} 2785
\item{[21]} Schutz B F 1972 {\it Ap. J. Supp.\/} {\bf 208} 319; {\bf 208} 343
\item{[22]} Moncrief V 1974 {\it Ann. Phys.\/} {\bf 88} 323
\item{[23]} Friedman J L and Schutz B F 1975 {\it Ap. J.\/} {\bf 200} 204
\item{[24]} Lund F 1973 {\it Phys. Rev.\/} D {\bf 8} 3253
\item{[25]} Demaret J and Moncrief V 1980 {\it Phys. Rev.\/} D {\bf 21} 2785
\item{[26]} Brown J D and Kucha\v r K V 1993 ``Dust as a standard of space
and time in canonical quantum gravity" in preparation
\item{[27]} Gibbons G W and Hawking S W 1977 {\it Phys. Rev.\/} D {\bf 15}
2752
\item{[28]} Brown J D and York J W 1993 {\it Phys. Rev.\/} D {\bf 47} 1420
\item{[29]} Clebsch A 1859 {\it J. Reine Angew. Math.\/} {\bf 56} 1
\item{[30]} van Dantzig D 1939 {\it Physica\/} {\bf 6} 693
\item{[31]} Taub A H 1959 {\it Arch. Ratl. Mech. Anal.\/} {\bf 3} 312
\item{[32]} Smarr L and York J W 1978 {\it Phys. Rev.\/} D {\bf 17} 2529
\item{[33]} Friedman J L 1978 {\it Commun. Math. Phys.\/} {\bf 62} 247
\item{[34]} Bekenstein J D 1987 {\it Ap. J.\/} {\bf 319} 207
\item{[35]} Arnowitt R, Deser S, and Misner C W 1962 The dynamics of general
relativity {\it Gravitation: An Introduction to Current Research\/} ed
L Witten (New York: Wiley and Sons)
\item{[36]} Dirac P A M 1967 {\it Lectures on Quantum Mechanics\/} (New York:
Academic Press)
\item{[37]} Ince E L 1956 {\it Ordinary Differential Equations\/} (New
York: Dover)
\item{[38]} Bretherton F P 1970 {\it J. Fluid Mech.\/} {\bf 44} 19
\item{[39]} Marsden J and Weinstein A 1983 {\it Physica\/} {\bf 7D} 305
\item{[40]} Carter B 1979 Perfect fluid and magnetic field conservation
laws in the theory of black hole accretion rings {\it Active Galactic
Nuclei\/} ed C Hazard and S Mitton (Cambridge: Cambridge University Press)
\item{[41]} K\"unzle H P and Nester J M 1984 {\it J. Math. Phys.\/} {\bf 25}
1009
%%%%%%%%%%%%%%%%%%%%%%%%%%%%%%%%%%%%%%%%%%%%%%%%%%%%%%%%%%%%%%%%%%%%%%%
\bye